\documentclass{aa}  
\usepackage{color}
\usepackage{graphicx}
\usepackage{amsmath}
\usepackage{mathrsfs}
\usepackage[varg]{txfonts}
\usepackage[T1]{fontenc}
\usepackage[utf8]{inputenc}
\usepackage{CJKutf8}
\usepackage{natbib,twoopt}
\usepackage{orcidlink}
\usepackage{hyperref} 
\usepackage{xcolor}
\usepackage{soul}
\usepackage{changes}
\definecolor{mylightblue}{rgb}{0.19,0.55,0.91}
\definecolor{coquelicot}{rgb}{1.00,0.22,0.00}
\hypersetup{colorlinks,breaklinks,
            urlcolor=mylightblue,citecolor=violet,
            anchorcolor=lightgray,linkcolor=mylightblue}
\bibpunct{(}{)}{;}{a}{}{,}             
\makeatletter
  \newcommandtwoopt{\citeads}[3][][]{\href{http://adsabs.harvard.edu/abs/#3}%
    {\def\hyper@linkstart##1##2{}%
     \let\hyper@linkend\@empty\citealp[#1][#2]{#3}}}
  \newcommandtwoopt{\citepads}[3][][]{\href{http://adsabs.harvard.edu/abs/#3}%
    {\def\hyper@linkstart##1##2{}%
     \let\hyper@linkend\@empty\citep[#1][#2]{#3}}}
  \newcommandtwoopt{\citetads}[3][][]{\href{http://adsabs.harvard.edu/abs/#3}%
    {\def\hyper@linkstart##1##2{}%
     \let\hyper@linkend\@empty\citet[#1][#2]{#3}}}
  \newcommandtwoopt{\citeyearads}[3][][]%
    {\href{http://adsabs.harvard.edu/abs/#3}
    {\def\hyper@linkstart##1##2{}%
     \let\hyper@linkend\@empty\citeyear[#1][#2]{#3}}}
\makeatother

\definecolor{blush}{rgb}{0.87, 0.36, 0.51}

\DeclareSymbolFont{mymath}{T1}{ybv}{m}{it}
\SetSymbolFont{mymath}{normal}{T1}{ybv}{m}{it}
\DeclareSymbolFontAlphabet{\mathnormal}{mymath}
\DeclareMathSymbol{\vel}{\mathalpha}{mymath}{`v}

\begin{document} 

   \title{SPONGE: Sulphur, Phosphorus, Oxygen, Nitrogen, and carbon Galactic Evolution}

   \subtitle{I. CNOPS elements as probes of physical processes in stars and galaxies}

   \author{D.~Romano
          \inst{1} \orcidlink{0000-0002-0845-6171},
          M.~Limongi
          \inst{2} \orcidlink{0000-0003-0636-7834},
          E.~Caffau
          \inst{3} \orcidlink{0000-0001-6011-6134},
          Y.~P.~Zou (\begin{CJK*}{UTF8}{gkai}邹益鹏\end{CJK*})
          \inst{4, 1} \orcidlink{0000-0002-5230-8010},
          P.~Bonifacio
          \inst{3} \orcidlink{0000-0002-1014-0635},
          L.~Monaco
          \inst{5} \orcidlink{0000-0002-3148-9836},
          L.~Colzi
          \inst{6} \orcidlink{0000-0001-8064-6394},
          E.~Spitoni
          \inst{7} \orcidlink{0000-0001-9715-5727},
          M.~Palla
          \inst{8} \orcidlink{0000-0002-3574-9578},
          S.~Tsiatsiou
          \inst{9} \orcidlink{0000-0003-2349-0316},
          D.~Nandal
          \inst{10} \orcidlink{0000-0003-1927-4397},
          J.~Alazzawi
          \inst{3} \orcidlink{0009-0000-0647-3675},
          A.~Falla
          \inst{11, 2} \orcidlink{0009-0005-0324-0637},
          F.~Fontani
          \inst{8, 12, 13} \orcidlink{0000-0003-0348-3418},
          A.~J.~Gallagher
          \inst{14} \orcidlink{0000-0003-3014-8981},
          R.~Hirschi
          \inst{15, 16} \orcidlink{0000-0001-8764-6522},
          F.~Lucertini
          \inst{17} \orcidlink{0000-0002-8785-0109},
          A.~Mucciarelli
          \inst{4, 1} \orcidlink{0000-0001-9158-8580},
          L.~Roberti
          \inst{18} \orcidlink{0000-0003-0390-8770},
          M.~Spite
          \inst{3} \orcidlink{0000-0002-2795-3421},\\
          and M.~Steffen
          \inst{19} \orcidlink{}
          }

   \institute{INAF, Osservatorio di Astrofisica e Scienza dello Spazio, Via Gobetti 93/3, 40129 Bologna, Italy\\
     \email{donatella.romano@inaf.it}
     \and INAF, Osservatorio Astronomico di Roma, Via Frascati 33, 00040 Monteporzio Catone, Italy
     \and LIRA, Observatoire de Paris, Universit\'e PSL, Sorbonne Universit\'e, Universit\'e Paris Cit\'e, CY Cergy Paris Universit\'e, CNRS, 92190 Meudon, France
     \and Dipartimento di Fisica e Astronomia, Universit\`a di Bologna, Via Gobetti 93/2, 40129 Bologna, Italy
     \and Universidad Andres Bello, Facultad de Ciencias Exactas, Departamento de Física y Astronomía, Instituto de Astrofísica, Autopista Concepción-Talcahuano, 7100 Talcahuano, Chile
     \and Centro de Astrobiolog{\'\i}a (CAB), CSIC-INTA, Ctra. de Ajalvir Km. 4, 28850 Torrej\'on de Ardoz, Madrid, Spain
     \and INAF, Osservatorio Astronomico di Trieste, Via Tiepolo 11, 34143 Trieste, Italy
     \and INAF, Osservatorio Astrofisico di Arcetri, Largo E. Fermi 5, 50125 Firenze, Italy
     \and Department of Astronomy, University of Geneva, Chemin Pegasi 51, 1290 Versoix, Switzerland
     \and Center for Astrophysics, Harvard and Smithsonian, 60 Garden Street, Cambridge, MA 02138, USA
     \and Dipartimento di Fisica, Sapienza Universit\`a di Roma, P.le Moro 5, 00185 Roma, Italy
     \and Max-Planck-Institut f\"{u}r extraterrestrische Physik, Giessenbachstra\ss e 1, 85748 Garching bei M\"{u}nchen, Germany
     \and LUX, Observatoire de Paris, PSL Research University, CNRS, Sorbonne Universit\'e, 92190 Meudon, France
     \and Institute for Applied Physics, Goethe University Frankfurt, Max-von-Laue-Straße 12, 60438 Frankfurt am Main, Germany
     \and Astrophysics Research Centre, Lennard-Jones Laboratories, Keele University, Keele ST5 5BG, UK
     \and Kavli IPMU (WPI), The University of Tokyo, 5-1-5 Kashiwanoha, Kashiwa 277-8583, Japan
     \and European Southern Observatory, Avenida Alonso de C{\'o}rdova 3107, Vitacura, Santiago, Chile
     \and INFN, Laboratori Nazionali del Sud, Via Santa Sofia 62, 95123 Catania, Italy
     \and Leibniz-Institut für Astrophysik (AIP), An der Sternwarte 16, 14482 Potsdam, Germany
             }

   \date{Received 16 November 2026 / Accepted 16 November 2026}

   \titlerunning{SPONGE I}
   \authorrunning{Romano et al.}

   \abstract{As fundamental bio-essential elements, carbon, nitrogen, oxygen, phosphorus, and sulphur (CNOPS) warrant a precise mapping of their cosmic origins. In this paper, we present the SPONGE project (Sulphur, Phosphorus, Oxygen, Nitrogen, and carbon Galactic Evolution), aimed at establishing a self-consistent galactic chemical evolution (GCE) framework capable of simultaneously reproducing local abundance benchmarks along with key observational constraints in external galaxies. Here, we focus on the Milky Way. We compile literature data and construct GCE models incorporating various single- and binary-star yields, taking stellar rotation and nova nucleosynthesis into account. Fast rotators are confirmed to be essential for primary $^{14}\mathrm{N}$ and $^{13}\mathrm{C}$ production at low metallicities. Although our homogeneous models do not reach the lowest $^{12}\mathrm{C}/^{13}\mathrm{C}$ ratios observed in some halo stars, this discrepancy should not be taken as a refutation of the model, as observational selection effects may apply. Predicted C and N isotopic ratios along the disc agree remarkably well with those inferred from radio observations of star-forming regions. Notably, our models explain Galactic P enrichment as originating predominantly from massive stars, without any ad hoc adjustment of the yields or the need for a significant nova contribution. Conversely, novae are required to synthesize non-negligible fractions of $^{13}\mathrm{C}$ and $^{15}\mathrm{N}$. Optimizing GCE trajectories against a comprehensive CNOPS ensemble simultaneously, rather than single diagnostics one at a time, establishes a robust baseline for future work. Remaining open questions---such as the predicted overproduction of $^{12}\mathrm{C}$ at low metallicities and $^{34}\mathrm{S}$ along the disc, alongside the abundance scatter of some key elements---provide concrete targets for subsequent papers in this series.}

   \maketitle
   \enlargethispage{-4.75pt} 
%

   \section{Introduction}
   \label{sec:intro}

The first generations of stars (Population~III stars, hereafter Pop~III) formed from a pristine mixture of hydrogen, helium, and trace amounts of lithium \citep{bromm2004,klessen2023}. Their evolution and ensuing deaths initiated galactic chemical enrichment, forging the first metals and seeding the interstellar medium (ISM) for subsequent stellar generations. Since then, each stellar generation has further enriched the gas through stellar winds and supernovae, connecting the physics of individual stars to the formation and evolution of galaxies and galaxy clusters \citep{tinsley1979,pagel1997,matteucci2021}.

Tracing this enrichment process requires empirical probes across cosmic epochs. By measuring elemental abundances in long-lived, low-mass stars, we can reconstruct the life cycles of local galaxies, placing tight constraints on their star-formation histories, galactic-scale outflows, and mass assembly via coronal gas accretion and satellite mergers. Complementary gas-phase abundance determinations grant access to elements lacking measurable stellar spectral features, although they reflect only a snapshot of chemical enrichment at the time of observation.

Carbon, nitrogen, oxygen, phosphorus, and sulphur (hereafter CNOPS) are fundamental to pre-biotic chemistry and the emergence of life as we know it. Multiple stellar generations are required to build a favourable chemical environment; tracking the evolution of CNOPS elements and their stable isotopes across cosmic time is therefore essential for both astrophysics and astrobiology. Addressing this question requires a multi-disciplinary approach combining empirical abundance measurements with detailed models of stellar and galactic chemical evolution (GCE). This is precisely the goal of the SPONGE (Sulphur, Phosphorus, Oxygen, Nitrogen, and carbon Galactic Evolution) collaboration.

Despite their crucial role, the Galactic evolution of CNOPS elements and their isotopes remains poorly constrained. In stars, CNO abundances can be derived from permitted and forbidden atomic lines, as well as molecular bands ($\mathrm{C}_2$, $\mathrm{CH}$, $\mathrm{NH}$, $\mathrm{OH}$, $\mathrm{CN}$, and $\mathrm{CO}$) in ultraviolet (UV), optical, and near-infrared (NIR) spectra \citep{barbuy2018,randich2021}. However, systematic differences between diagnostic methods yield conflicting abundances, even for the Sun \citep{caffau2009,amarsi2019,mashonkina2024}. Large spectroscopic surveys have provided $^{12}\mathrm{C}$, $^{14}\mathrm{N}$, and $^{16}\mathrm{O}$ abundances for thousands of Galactic stars \citep[e.g.][]{buder2019,franchini2020,franchini2021,kisku2021,li2024}, but these samples are largely restricted to giants, where atmospheric compositions are altered by evolutionary mixing processes. In contrast, the atmospheres of dwarf and unmixed subgiant stars retain their birth composition, making them excellent tracers of chemical enrichment on a Galactic scale.

Phosphorus abundance determinations in stars are hindered by the element's low cosmic abundance and scarce, weak optical lines \citep[see e.g.][]{takeda2024,aschenbrenner2025}, forcing studies to rely predominantly on NIR diagnostics \citep{caffau2011,barbuy2025}. Sulphur is similarly probed via NIR transitions, optical features---notably Multiplet~8 near 6757\,\AA, which forms close to local thermodynamic equilibrium (LTE) conditions \citep{caffau2010,lucertini2023}---and space-based UV lines \citep{roederer2014,roederer2016}.

Mapping chemical enrichment patterns across the entire Galactic disc requires probes such as classical Cepheids, open clusters, \ion{H}{ii} regions, and molecular clouds \citep[e.g.][]{wilson1994,milam2005,arellanocordova2020,colzi2022,magrini2023,yan2023,bono2024}. Gas-phase tracers present distinct challenges. Molecular isotopologue ratios can be biased relative to true isotopic ratios by local physical processes such as isotope-selective photodissociation and chemical fractionation \citep[e.g.][]{visser2009,roueff2015,colzi2020,nomura2023}, which depend on the physical properties of the molecular cloud, such as density, temperature, or exposure to an external UV field. Similarly, abundances derived from the emission spectra of photoionized regions often rely on simplified assumptions that neglect thermal and spatial stratification \citep[e.g.][]{marconi2024} or dust depletion of refractory and semi-refractory species \citep[e.g.][]{konstantopoulou2024,psaradaki2024}. Nevertheless, these tracers are indispensable for probing regions where stellar spectroscopy is not feasible, including high-redshift emitters.

As the first paper in the SPONGE series, this study establishes a baseline for CNOPS evolution in galaxies by critically evaluating GCE model predictions for the Milky Way against state-of-the-art observational data, highlighting key discrepancies and missing ingredients in current nucleosynthesis prescriptions. Section~\ref{sec:model} outlines the assumptions of the adopted homogeneous chemical evolution model. Section~\ref{sec:data} describes the observational dataset, followed by a comparison with model predictions in Sect.~\ref{sec:res}. Our findings are discussed in Sect.~\ref{sec:disc}, where we also draw our main conclusions. Extensions to other galaxies and high-redshift environments are deferred to future papers.

   \section{Chemical evolution model}
   \label{sec:model}

We employ a revised version of the classic two-infall chemical evolution model for the Milky Way originally proposed by \citet{chiappini1997,chiappini2001}. Specifically, we adopt the updated framework of \citet{spitoni2019} and subsequent refinements by \citet{palla2020} and \citet{molero2023}. The foundational premise of the two-infall scenario is that the Galaxy grows via two main gas accretion events. The initial episode drives the rapid collapse and formation of the in situ halo and chemically defined thick disc. A second, delayed accretion phase governs the gradual assembly of the thin disc. The mathematical framework and basic assumptions underlying our GCE model are detailed in Appendix~\ref{app:basic}, whereas the adopted stellar nucleosynthesis prescriptions are described below.

    \subsection{Stellar nucleosynthesis prescriptions}
    \label{sec:nuc}

To track the evolution of CNOPS elements and their isotopes, we account for four primary nucleosynthetic sources: low- and intermediate-mass stars (LIMS), massive stars, Type~Ia supernovae (SNe~Ia), and classical novae, for which we explore different nucleosynthesis prescriptions.

    \subsubsection{Low- and intermediate-mass stars}
    \label{sec:lmnuc}
    
LIMS span the initial mass range $1 \le m/M_\odot \le 8$. Our baseline yield set relies on non-rotating stellar evolutionary models coupled to a full nuclear network including isotopes up to the end of the $s$-process path, available from the Full-network Repository of Updated Isotopic Tables and Yields (FRUITY\footnote{\url{http://fruity.oa-abruzzo.inaf.it/}}) database \citep{cristallo2011,cristallo2015}. This grid encompasses initial masses from $1.3$ to $6\,M_\odot$ computed across 13 initial metallicities ranging from $Z = 4.8 \times 10^{-5}$ to $3 \times 10^{-2}$ (the lowest-metallicity tracks are explicitly $\alpha$-enhanced). We select models computed under standard assumptions regarding the efficiency and spatial extent of the internal $^{13}\mathrm{C}$ pocket that forms in the core-envelope transition region \citep{cristallo2009}.

As a second alternative, we adopt the yields calculated with the \texttt{ATON} stellar evolution code \citep{ventura1998,ventura2013,ventura2014,ventura2018,ventura2020,ventura2021}. This set covers the full $1$--$8\,M_\odot$ range across nine initial metallicities (from $Z = 3 \times 10^{-7}$ to $Z = 0.04$). The integrated nuclear network is rather limited but sufficient to follow in detail the evolution of CNO elements.

The third yield set adopted in this study is based on the Monash stellar evolution code \citep{lugaro2012,fishlock2014,karakas2016,karakas2018}. It covers the metallicity range $1 \times 10^{-4} \le Z \le 3 \times 10^{-2}$ and spans initial masses ranging from $0.9$--$1\,M_\odot$ to $6$--$8\,M_\odot$, depending on the specific metallicity grid. For this set as well, we adopt the yields computed under standard assumptions regarding the size of the $^{13}\mathrm{C}$ pocket. Furthermore, we select models with overshooting and $\alpha$-enhancement when available.

    \subsubsection{High-mass stars}
    \label{sec:hmnuc}

For massive stars ($m > 10\,M_\odot$), our baseline yields rely on computations performed with the Frascati RAphson Newton Evolutionary Code \citep[\texttt{FRANEC};][]{CL13,limongi2018,roberti2024}. All models have been followed from the pre-main sequence up to the onset of iron core collapse. An extended nuclear network, including 335 isotopes from neutrons to $^{209}\mathrm{Bi}$, is fully coupled to both the structural evolution and the chemical mixing induced by various physical processes (such as convection, semi-convection, diffusion, and rotationally driven mixing). Mass loss is accounted for using stage-dependent prescriptions. Rotation is included according to the shellular rotation approach \citep{MM97}. The angular momentum transport due to shear instabilities and meridional circulation is treated by solving an advection-diffusion equation \citep{MM00}, properly coupling rotation and mass loss. The stellar explosions, along with the associated explosive nucleosynthesis and remnant masses, are computed using the HYdrodynamic Ppm Explosion with Radiation diffusION (\texttt{HYPERION}) code \citep{LC20}, which solves the hydrodynamic equations and radiation transport in the flux-limited diffusion approximation. The same nuclear network adopted in the hydrostatic (\texttt{FRANEC}) calculations is coupled to these equations to track the abundances of all considered nuclear species in a self-consistent way. The yield tables can be downloaded from the Online Repository of the Franec Evolutionary Output (O.R.F.E.O.\footnote{\url{https://orfeo.oa-roma.inaf.it/}}).

We implement two distinct configurations, designated as Set~M and Set~R. Both account for initial equatorial velocities of $v_{\mathrm{ini}} = 0$, $150$, and $300\,\mathrm{km\,s^{-1}}$\footnote{The ratio between the angular and the critical ($v_{\mathrm{crit}}$) velocities at the beginning of the main sequence varies as a function of stellar mass and metallicity \citep[see][their figure~4]{limongi2018}.}, cover the mass range $13$--$120\,M_\odot$, and are computed across initial metallicities $[\mathrm{Fe}/\mathrm{H}] = -3$, $-2$, $-1$, $0$, and $0.3$. Set~M utilizes the mixing and fallback scheme of \citet{umeda2002}, where the inner boundary of the mixed region is determined by enforcing $[\mathrm{Ni}/\mathrm{Fe}] = 0.2$, the outer boundary is fixed at the base of the oxygen-burning shell, and the mass cut is set by assuming the ejection of $0.07\,M_\odot$ of $^{56}\mathrm{Ni}$. Set~R is computed in the same way as Set~M; in this case, however, only stars in the mass range $13 \le m/M_\odot \le 25$ are assumed to successfully explode as core-collapse supernovae (CCSNe), while more massive stars collapse directly to black holes, thereby enriching the ISM solely via pre-explosive stellar winds.

As a further reference, we adopt the yields reviewed by \citet{nomoto2013}, which account for stellar mass loss but omit rotational mixing. The stellar models cover the $13$--$40\,M_\odot$ mass range (except for the Pop~III grid, where a higher upper mass limit is considered) across initial metallicities ranging from pristine ($Z = 0$) to super-solar ($Z = 0.05$). For $m \ge 20\,M_\odot$, the authors provide prescriptions for both standard CCSNe and hyper-energetic supernovae (hypernovae, HNe). In our GCE models, we implement a 50:50 mixture of CCSNe and HNe, independent of metallicity. Models with different mixtures are available upon request.

Finally, we consider a set of yields obtained from massive star models computed with the Geneva Stellar Evolution Code (\texttt{GENEC}\footnote{\url{https://github.com/GESEG/GENEC}}) that span metallicities from $Z = 0$ to $Z = 0.020$ (Nandal et al., in prep.). This grid covers initial masses $9 \le m/M_\odot \le 120$ with either zero or moderate rotation ($v_{\mathrm{ini}} = 0.4\,v_{\mathrm{crit}}$), except for the Pop~III models of \citet{tsiatsiou2024}, which feature rapid rotation ($v_{\mathrm{ini}} = 0.7\,v_{\mathrm{crit}}$). Because the evolution of these models has been followed only up to core O depletion, the yields of all the isotopes synthesized during the more advanced burning stages or during the explosion cannot be reliably determined. For this reason, we restrict our use of the \texttt{GENEC} set to isotopes up to $^{22}\mathrm{Ne}$. The sole exception is iron, for which we adopt the following rules: stars with $m = 12\,M_\odot$ eject $0.04\,M_\odot$ of $^{56}\mathrm{Fe}$. For stars with $m \ge 15\,M_\odot$, the ejected $^{56}\mathrm{Fe}$ mass is set to $0.07\,M_\odot$ for models adopting the \citet{maeder1992} remnant-mass prescriptions, whereas under the \citet{patton2020} remnant-mass scheme, this yield is adopted only for models with $m_{\mathrm{CO}} < 12\,M_\odot$ and $m_{\mathrm{rem}} < 5\,M_\odot$; all other models eject no $^{56}\mathrm{Fe}$.

For all yield configurations, linear interpolation and extrapolation are applied when required progenitor masses or metallicities fall between or beyond the tabulated grid points, following standard practice in GCE modelling.

    \subsubsection{Binary stars}

For SNe~Ia, we assume a single-degenerate evolutionary scenario \citep{matteucci1986,matteucci2001}. The progenitor system consists of a close binary in which a white dwarf accretes material from a less-evolved companion. Upon approaching the Chandrasekhar mass limit, the white dwarf undergoes a thermonuclear explosion via carbon deflagration. The chemical yields for this channel are taken from the classic W7 model of \citet{iwamoto1999}.

Finally, we incorporate chemical enrichment from classical novae, which represent crucial sites for the synthesis of CNO isotopes \citep{romano2003,romano2017} and possibly phosphorus \citep{bekki2024}. The adopted nova yields are based on the ONe4 model of \citet{jose1998}, with empirical modifications to the $^{13}\mathrm{C}$, $^{15}\mathrm{N}$, or $^{18}\mathrm{O}$ yields to reproduce observations (see Sect.~\ref{sec:mix}).

   \section{Observational data}
   \label{sec:data}
   
\subsection{Local tracers: solar-neighbourhood low-mass stars}
\label{sec:data_SN}

To constrain the local evolution of carbon, nitrogen, and oxygen, we compile high-resolution spectroscopic data across a wide range of metallicities. Our primary sample consists of 714 dwarf and subgiant stars analysed by \citet{bensby2026}; their CNO abundances include corrections for departures from LTE conditions. We add to this sample C and O abundances for 187 F and G dwarfs from \citet{amarsi2019}, who derived 3D non-LTE abundances from high-resolution HARPS, FEROS, UVES, and FIES spectra; to ensure consistency with our solar neighbourhood GCE tracks, we remove 35 stars identified as part of the accreted low-$\alpha$ halo population. At intermediate metallicities, $-1.5 \le [\mathrm{Fe}/\mathrm{H}] < 0$, we complement our dataset with unevolved stars from \citet{israelian2004}, who measured N and O abundances via spectral synthesis of near-UV $\mathrm{NH}$ and $\mathrm{OH}$ molecular bands. At the metal-rich end ($[\mathrm{Fe}/\mathrm{H}] \gtrsim -0.25$), we include 74 solar-type stars observed with UVES \citep{suarezandres2016} alongside the HARPS solar-twin sample of \citet{botelho2020}.

During the ascent of the red giant branch (RGB), stellar surface abundances are altered by deep mixing. Therefore, when incorporating halo giant samples, we restrict our selection to unmixed stars \citep[following the definition of][]{spite2005}. Specifically, we adopt the lower RGB subset from \citet[][with C and O abundances from \citealt{cayrel2004}]{spite2005}, lower RGB stars from \citet[][their table~1]{rossi2024}, and CNO abundances of the old subgiants HD\,140283 \citep{spite2021} and CES\,0422$-$3715 \citep{fernandesdemelo2024}.

To constrain C and O isotopic ratios, we combine optical molecular line diagnostics with high-resolution NIR ($K$- and $M$-band) spectroscopy. In the low-metallicity domain, optical $^{12}\mathrm{C}/^{13}\mathrm{C}$ ratios of low-mass giants and dwarfs are drawn from \citet{molaro2023}, alongside the benchmark measurement of $^{12}\mathrm{C}/^{13}\mathrm{C} = 33_{-6}^{+12}$ in the unmixed subgiant HD\,140283 \citep{spite2021}. At solar metallicities, $-0.5 \le [\mathrm{Fe}/\mathrm{H}] < 0.5$, we adopt $^{12}\mathrm{C}/^{13}\mathrm{C}$ ratios of solar twins from \citet{botelho2020}, together with $^{12}\mathrm{C}/^{13}\mathrm{C}$ and $^{16}\mathrm{O}/^{18}\mathrm{O}$ determinations in fully convective M dwarfs observed with CFHT/SPIRou \citep{gonzalezpicos2025}, Keck~II/KPIC \citep{wang2026}, and IRTF/iSHELL \citep{crossfield2019}. Note that neither $^{15}\mathrm{N}$ nor $^{17}\mathrm{O}$ abundances are measured in unevolved stars.

Unlike C and N, photospheric abundances of P and S are unaffected by evolutionary mixing as stars climb the giant branches, making giants excellent tracers of P and S Galactic evolution. Data for phosphorus are drawn mainly from high-resolution NIR spectroscopy of atomic \ion{P}{i} features. \citet{caffau2011} derived P abundances for 20 cool disc stars observed with VLT/CRIRES using \ion{P}{i} Multiplet~1 at $10\,500$--$10\,820$\,\AA, while \citet{caffau2019} targeted \ion{P}{i} Multiplet~4 lines across 40 disc stars using TNG/GIANO. The sample around solar metallicity is expanded by including the data from \citet{jian2026}, who measured P abundances from TNG/GIANO-B NIR spectra of 102 young to intermediate-age giant stars (82 stars in 24 open clusters and 20 Cepheids), confirming the declining $[\mathrm{P}/\mathrm{Fe}]$ trend towards higher metallicities found in previous studies. To extend our baseline to lower-metallicity regimes, we adopt abundances from \citet{roederer2014} derived from HST/STIS near-UV spectra (primarily the \ion{P}{i} $\lambda\,2136.18$\,\AA\ line) for 14 stars spanning $-3.8 < [\mathrm{Fe}/\mathrm{H}] < -0.1$. After applying their optical-to-UV scale offset correction, their halo sample displays a constant plateau of $\langle [\mathrm{P}/\mathrm{Fe}] \rangle = +0.04 \pm 0.10$, indicating primary production of P in massive stars. Finally, we include the large, homogeneous sample of 750 FGK stars by \citet{gakis2026}.

For sulphur, we compile spectroscopic datasets covering metallicities from the extremely metal-poor halo to the local thin and thick discs. For $[\mathrm{Fe}/\mathrm{H}] < -2$, we adopt the VLT/UVES 3D non-LTE measurements of \citet{spite2011}, showing that $[\mathrm{S}/\mathrm{Fe}]$ exhibits an $\alpha$-element plateau at very low metallicities. Across the broad metallicity interval $-2.0 \le [\mathrm{Fe}/\mathrm{H}] \le +0.5$, we take advantage of the large samples of \citet[][LTE abundances]{perdigon2021} and \citet[][non-LTE abundances]{bensby2026}. Finally, we include in our analysis the high-precision non-LTE S determinations from \citet{lucertini2022} for kinematically selected disc stars (21 thick-disc and 30 thin-disc stars), based on spectral synthesis of \ion{S}{i} Multiplets~1, 6, and 8.

Sulphur isotopic ratios provide unique constraints on GCE and stellar models. Unlike CNO isotopes, which are primarily synthesized via CNO cycling and helium burning \citep[][and references therein]{romano2022}, $^{32}\mathrm{S}$ and $^{34}\mathrm{S}$---the two most abundant stable isotopes, representing 95 and 4.2 per cent of solar sulphur, respectively---are mainly produced during hydrostatic and explosive O- and Ne-burning phases in CCSNe and SNe~Ia \citep{woosley1995,keegans2023}. In recent years, high-resolution sub-millimetre and millimetre spectroscopy targeting rotational transitions of S-bearing species ($\mathrm{SO}$, $\mathrm{SO}_2$, $\mathrm{CS}$, and $\mathrm{H}_2\mathrm{S}$) and their isotopologues has enabled precise determinations of the $^{32}\mathrm{S}/^{34}\mathrm{S}$ ratio in the circumstellar envelopes of asymptotic giant branch (AGB) stars. Measurements across O-rich, S-type, and C-rich AGB stars generally cluster around the solar ratio \citep[e.g.][]{unnikrishnan2024,wallstroem2024,gorai2025}. Crucially, stellar evolution models demonstrate that the $^{32}\mathrm{S}/^{34}\mathrm{S}$ ratio remains unaltered by internal processes throughout the evolution of LIMS (A.~Karakas, priv.~comm.), making these circumstellar envelope measurements robust empirical benchmarks for GCE models.

Finally, whenever possible, we derive homogeneous kinematics and ages for the compiled stellar sample, following the procedures detailed in Appendices~\ref{app:kin} and~\ref{app:ages}, respectively.

\subsection{Radial gradients}
\label{sec:data_DISC}

We complement our local baseline with radial abundance profiles of C, N, O, S, and their stable isotopes. We adopt LTE abundances of C, O, and S from high-resolution UVES, HARPS-N, and ESPaDOnS spectra of 245 classical Cepheids from the C-MetaLL survey \citep{trentin2023,trentin2024}, as well as non-LTE O and S abundances for 401 Galactic Cepheids spanning Galactocentric distances from $5$ to $29\,\mathrm{kpc}$ \citep{nunnari2026}. Furthermore, we adopt the C, N, O, and S radial gradients of \citet{mendezdelgado2022}, derived from deep optical spectra of 42 \ion{H}{ii} regions within the temperature-fluctuation paradigm, incorporating updated \textit{Gaia} EDR3 distances.

As for carbon isotopic ratios, measurements across multiple molecular species consistently demonstrate a positive radial gradient \citep{colzi2026}. We adopt $^{12}\mathrm{C}/^{13}\mathrm{C}$ ratios from $\mathrm{CN}$ \citep{savage2002,milam2005}, $\mathrm{C}^{18}\mathrm{O}$ \citep{langer1990,wouterloot1996,keene1998}, $\mathrm{H}_2\mathrm{CO}$ \citep{henkel1980,henkel1982,henkel1983,henkel1985,yan2019}, $\mathrm{CH}^+$ \citep{ritchey2011}, $\mathrm{CH}$ \citep{jacob2020}, $\mathrm{CS}$ \citep{yan2023}, and $\mathrm{HCO}^+$ \citep{luo2024}. Knowledge of the $^{12}\mathrm{C}/^{13}\mathrm{C}$ gradient is essential to infer other key isotopic ratios from radio observations: $^{16}\mathrm{O}/^{18}\mathrm{O}$ \citep{wilson1994,wouterloot2008}, $^{14}\mathrm{N}/^{15}\mathrm{N}$ \citep{colzi2018,colzi2022}, and $^{32}\mathrm{S}/^{34}\mathrm{S}$ \citep{yu2020,yan2023}. Direct determinations of $^{16}\mathrm{O}/^{18}\mathrm{O}$ from $\mathrm{OH}$ \citep{polehampton2005} and $^{32}\mathrm{S}/^{34}\mathrm{S}$ from $^{13}\mathrm{CS}$ \citep{yan2023} are also available and are adopted here, although sample sizes remain small due to the low detection rates of the rare isotopologues. Moreover, optical depth effects in the observed transitions can introduce additional uncertainties in the derived isotopic ratios, particularly when the assumption of optically thin emission is not fully justified. Finally, measurements of $^{18}\mathrm{O}/^{17}\mathrm{O}$ \citep{zou2023} and $^{34}\mathrm{S}/^{33}\mathrm{S}$ \citep{yu2020,yan2023} in molecular clouds provide crucial constraints on the synthesis of $^{17}\mathrm{O}$ and $^{33}\mathrm{S}$, as stellar determinations are either restricted to giant stars susceptible to evolutionary mixing \citep[e.g.][]{denutte2017} or absent. A full discussion of $^{17}\mathrm{O}$ and $^{33}\mathrm{S}$ evolution is deferred to Paper~II in this series (Zou et al., in prep.).

    \begin{figure*}
    \centering
    \includegraphics[width=0.49\textwidth]{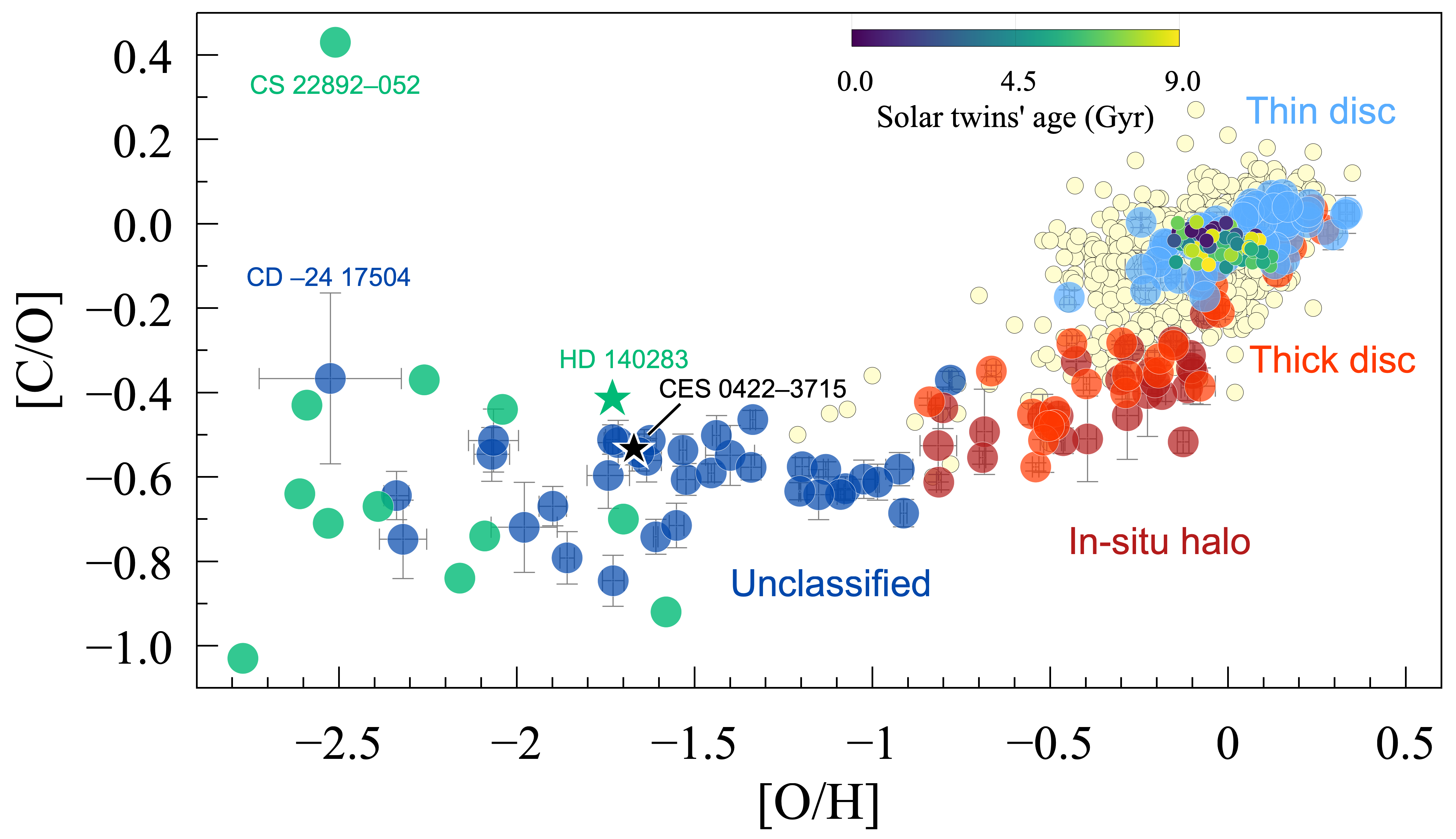}
    \includegraphics[width=0.49\textwidth]{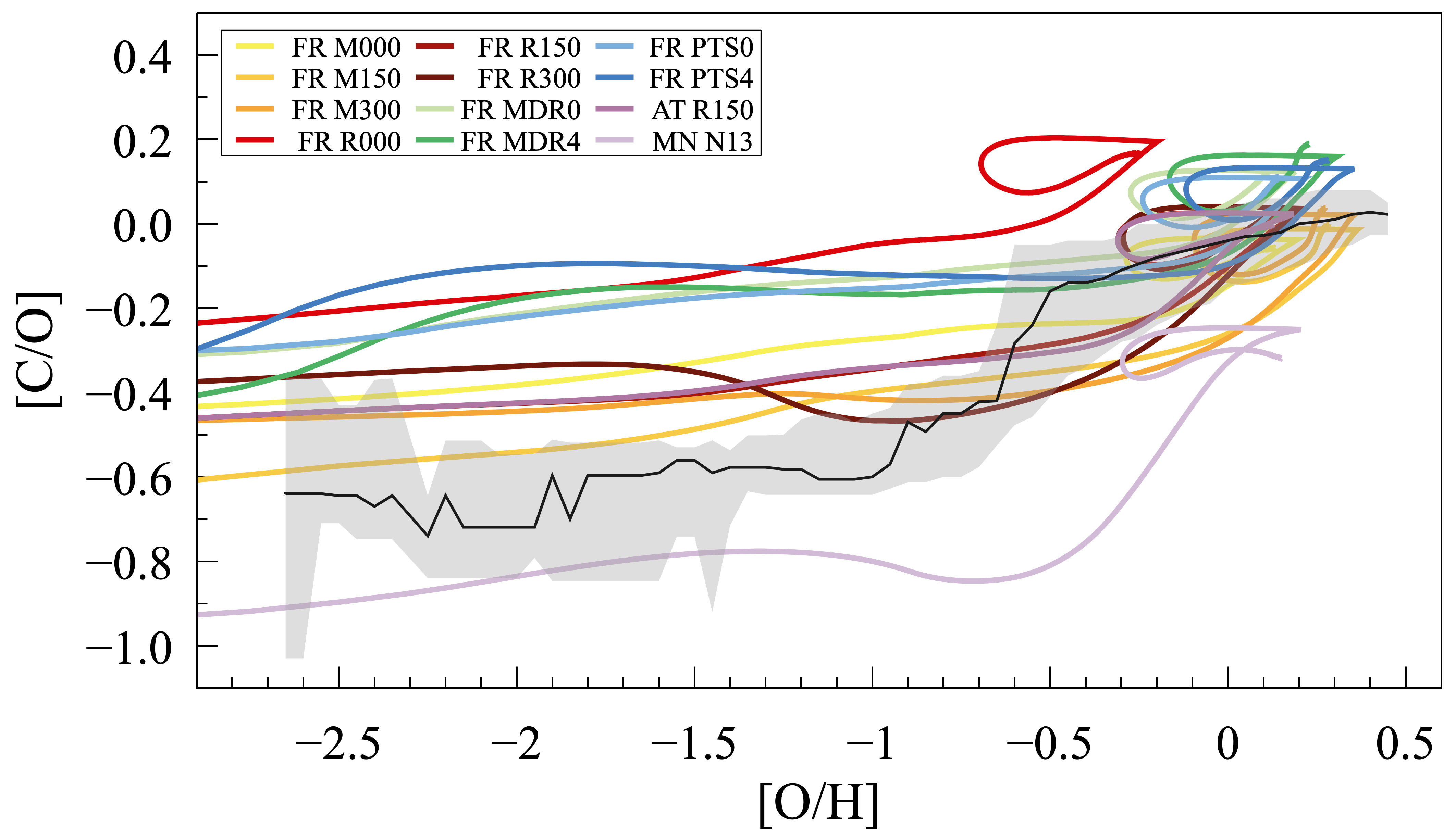}
    \caption{\textit{Left:} $[\mathrm{C}/\mathrm{O}]$ vs. $[\mathrm{O}/\mathrm{H}]$ diagram for late-type stars in the solar neighbourhood. Light-blue, orange, red, and blue circles denote 3D non-LTE abundances for thin-disc, thick-disc, in-situ halo, and unclassified F--G dwarfs from \citet{amarsi2019}. Green circles represent unmixed lower-RGB stars from the `First Stars' programme \citep{cayrel2004,spite2005}. Star-shaped markers highlight the Methuselah star \citep{spite2021} and the unmixed halo giant CES\,0422$-$3715 \citep{fernandesdemelo2024}. Cream circles are non-LTE abundances of dwarfs and subgiants from \citet{bensby2026}. Small age-coded circles indicate solar twins \citep{botelho2020}. \textit{Right:} theoretical $[\mathrm{C}/\mathrm{O}]$ vs. $[\mathrm{O}/\mathrm{H}]$ tracks for various yield prescriptions (thick coloured lines; see text). The median observational trend (thin black line) and 16$^\mathrm{th}$--84$^\mathrm{th}$ percentile dispersion envelope (grey band) are computed using a sliding boxcar window of width $\Delta[\mathrm{O}/\mathrm{H}] = 0.35\,\mathrm{dex}$ with a minimum threshold of 6 stars per bin.}
    \label{fig:CO_OH}
    \end{figure*}
        
    \begin{figure*}
    \centering
    \includegraphics[width=\columnwidth]{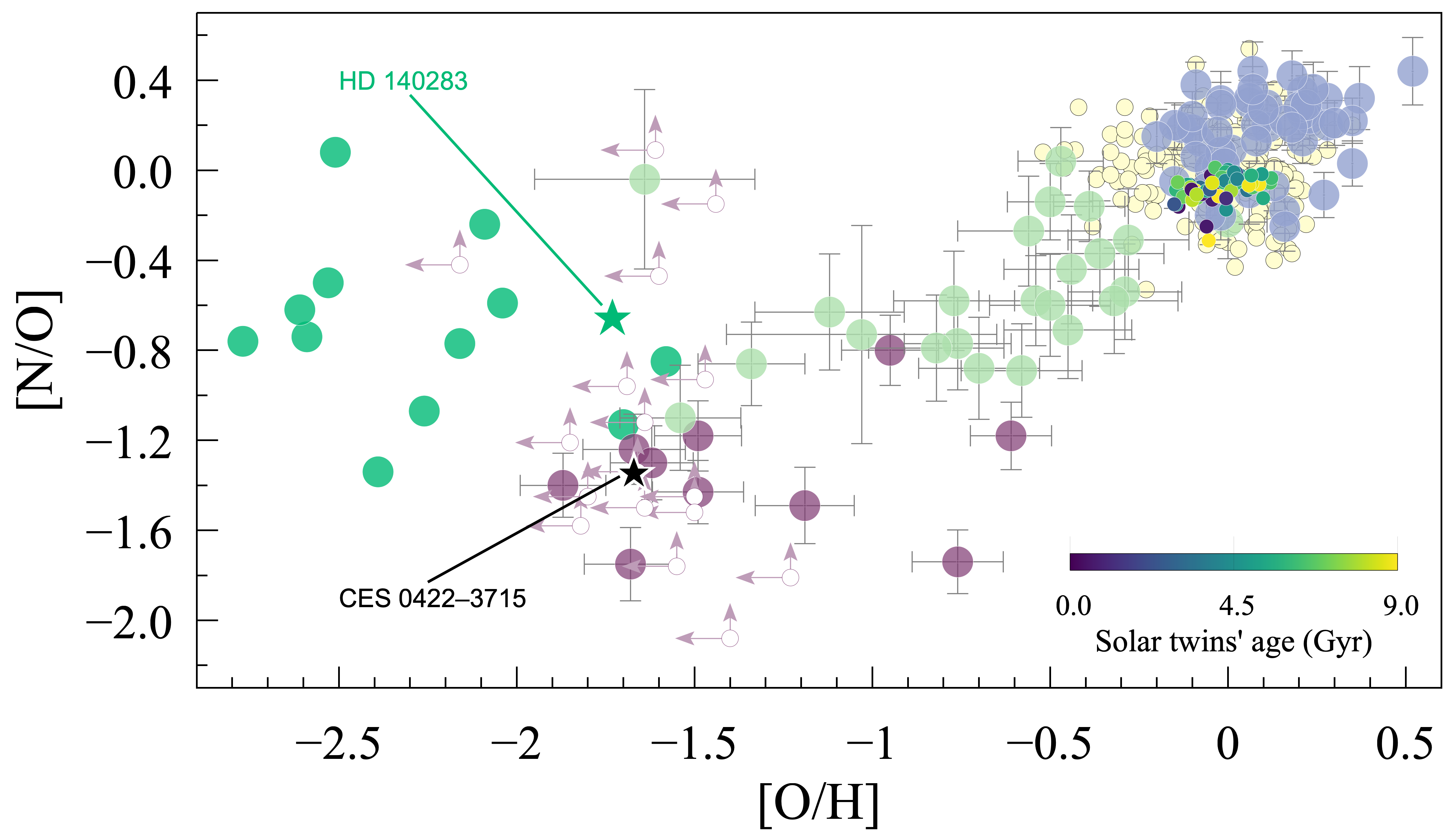}
    \includegraphics[width=\columnwidth]{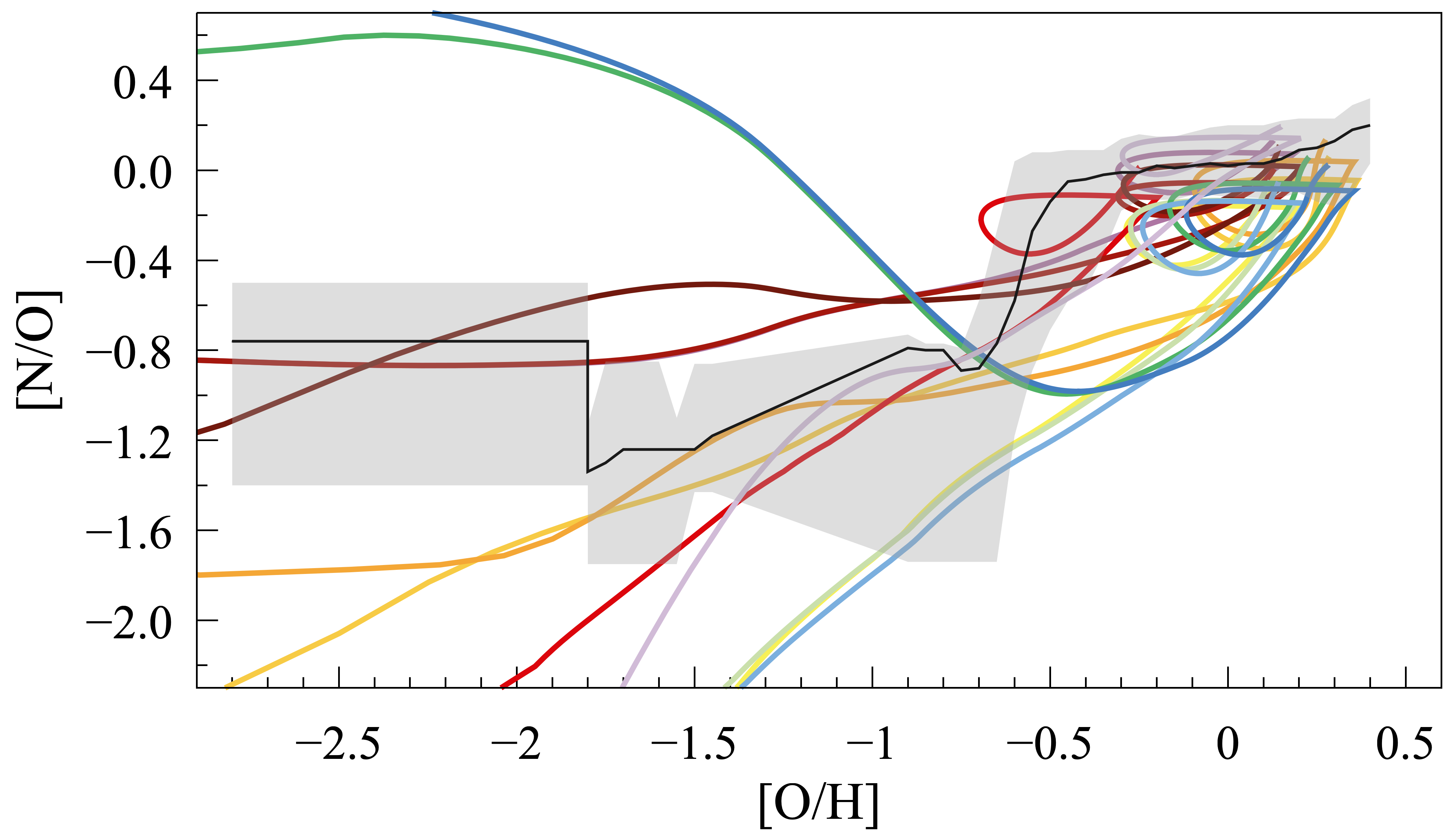}
    \caption{\textit{Left:} same as Fig.~\ref{fig:CO_OH}, but for $[\mathrm{N}/\mathrm{O}]$ vs. $[\mathrm{O}/\mathrm{H}]$. Shared datasets \citep{cayrel2004,spite2005,spite2021,botelho2020,fernandesdemelo2024,bensby2026} are as described in Fig.~\ref{fig:CO_OH}. Additional data include lower-RGB halo stars from \citet[][dark magenta filled and open circles for measurements and limits, respectively]{rossi2024}, unevolved stars from \citet[][light green circles]{israelian2004}, and solar-type stars from \citet[][light steel-blue circles]{suarezandres2016}. \textit{Right:} model predictions and observational median trend defined as in Fig.~\ref{fig:CO_OH}. For $[\mathrm{O}/\mathrm{H}] < -1.8$, data are binned into a single interval to account for sample sparseness.}
    \label{fig:NO_OH}
    \end{figure*}
        
Throughout this paper, we adopt as a reference the solar abundance pattern of \citet{asplund2021}. According to these authors, the protosolar mass fractions of hydrogen, helium, and metals are $0.7121$, $0.2725$, and $0.0154$, respectively. The photospheric CNOPS abundances, $A(\mathrm{C}) = 8.46 \pm 0.04$, $A(\mathrm{N}) = 7.83 \pm 0.07$, $A(\mathrm{O}) = 8.69 \pm 0.04$, $A(\mathrm{P}) = 5.41 \pm 0.03$, and $A(\mathrm{S}) = 7.12 \pm 0.03$, agree with those determined by \citet{caffau2011SoPh}, $A(\mathrm{C}) = 8.50 \pm 0.06$, $A(\mathrm{N}) = 7.86 \pm 0.12$, $A(\mathrm{O}) = 8.76 \pm 0.07$, $A(\mathrm{P}) = 5.46 \pm 0.04$, and $A(\mathrm{S}) = 7.16 \pm 0.05$, within the respective errors. For the isotopic ratios, whenever possible we adopt the values measured in the solar photosphere \citep[$^{12}\mathrm{C}/^{13}\mathrm{C} = 91.4 \pm 1.3$, $^{16}\mathrm{O}/^{18}\mathrm{O} = 511 \pm 10$, and $^{16}\mathrm{O}/^{17}\mathrm{O} = 2738 \pm 118$;][]{ayres2013}; otherwise, we adopt estimates for the solar wind \citep[$^{14}\mathrm{N}/^{15}\mathrm{N} = 441 \pm 6$;][]{marty2011} or IAEA-S-1, a replacement standard for the Canyon Diablo meteorite \citep[$^{32}\mathrm{S}/^{34}\mathrm{S} \simeq 22.6$;][]{ding1999}.

   \section{Results}
   \label{sec:res}

\subsection{Solar neighbourhood}
\label{sec:mono}

In Fig.~\ref{fig:CO_OH}, we present the observational data compilation (left panel) alongside our GCE model trajectories (right panel) for $[\mathrm{C}/\mathrm{O}]$ as a function of $[\mathrm{O}/\mathrm{H}]$ in the solar neighbourhood. The characteristic back-and-forth excursions in the theoretical tracks arise from our double-infall assumption, whereby the accretion of primordial gas at the onset of the second infall episode temporarily dilutes the ISM, moving the tracks towards lower metallicity before subsequent star formation resumes the enrichment towards higher metallicities. The observational trend displays two distinct regimes. At low metallicities, $[\mathrm{O}/\mathrm{H}] \lesssim -1.0$, the data exhibit a relatively flat baseline centred around $[\mathrm{C}/\mathrm{O}] \approx -0.6$, albeit with significant scatter at the lowest metallicities ($[\mathrm{O}/\mathrm{H}] < -2.0$). At higher metallicities, $[\mathrm{C}/\mathrm{O}]$ undergoes a steep rise towards nearly solar values, followed by a smoother increase in the thin-disc population. Thick-disc stars systematically occupy a lower $[\mathrm{C}/\mathrm{O}]$ sequence compared to thin-disc stars of the same $[\mathrm{O}/\mathrm{H}]$, reflecting the delayed injection of carbon from low-mass stars during the slower assembly of the thin disc. To evaluate the impact of stellar physics on this evolution, the right-hand panel of Fig.~\ref{fig:CO_OH} compares the median empirical trend against GCE tracks computed with different combinations of yields for LIMS and massive stars. The model abbreviations in the legend denote the adopted nucleosynthesis prescriptions as follows:
\begin{itemize}
    \item LIMS yields: FR corresponds to the FRUITY baseline grid \citep{cristallo2011,cristallo2015}, AT to \texttt{ATON} yields \citep{ventura2013,ventura2014,ventura2018,ventura2020,ventura2021}, and MN to Monash yields \citep{lugaro2012,fishlock2014,karakas2016,karakas2018}.
    \item Massive star yields: M000, M150, and M300 denote the \citet{limongi2018} and \citet{roberti2024} Set~M with initial rotation velocities $v_{\mathrm{ini}} = 0$, $150$, and $300\,\mathrm{km\,s^{-1}}$, respectively; R000, R150, and R300 similarly correspond to Set~R. MDR0 and MDR4 adopt \texttt{GENEC} yields (Nandal et al., in prep.) with $v_{\mathrm{ini}} = 0$ and $v_{\mathrm{ini}} = 0.4\,v_{\mathrm{crit}}$ using the \citet{maeder1992} remnant-mass prescription; PTS0 and PTS4 adopt \texttt{GENEC} yields with the \citet{patton2020} remnant-mass prescription. Lastly, N13 denotes the \citet{nomoto2013} yields.
\end{itemize}

In the early phases ($[\mathrm{O}/\mathrm{H}] \lesssim -1.0$), the $[\mathrm{C}/\mathrm{O}]$ ratio is governed by massive star yields, with model predictions spanning a broad range depending on the implementation of stellar rotation, mass-loss schemes, and remnant-mass cuts. Comparing the results obtained with \citeauthor{limongi2018}'s (\citeyear{limongi2018}) Set~R and Set~M, for a given initial rotational velocity, models using Set~R predict higher $[\mathrm{C}/\mathrm{O}]$ ratios than those using Set~M. This difference is mainly due to the fate of the most massive stars: in Set~R, progenitors more massive than $25\,M_\odot$ undergo direct collapse to a black hole without exploding. As a result, their oxygen-rich mantles remain trapped inside the black hole remnant rather than being ejected, drastically reducing their contribution to the oxygen enrichment of the ISM compared to Set~M. Conversely, for a given set, models incorporating stellar rotation predict lower $[\mathrm{C}/\mathrm{O}]$ ratios (by about $0.2\,\mathrm{dex}$) relative to non-rotating ones. This reduction arises from rotation-induced mixing, which continuously transports fresh helium into the burning core during core He burning. This favours the depletion of $^{12}\mathrm{C}$ via the $^{12}\mathrm{C}(\alpha, \gamma)^{16}\mathrm{O}$ reaction, effectively shifting the chemical yields towards higher oxygen production at the expense of carbon.

The upturn in $[\mathrm{C}/\mathrm{O}]$ at $[\mathrm{O}/\mathrm{H}] > -0.5$ marks the onset of chemical enrichment from low-mass stars ($m \le 3\,M_\odot$). As these stars enter the AGB phase, repeated third dredge-up episodes turn them into carbon stars; coupled with efficient mass loss, this leads to the injection of large quantities of primary $^{12}\mathrm{C}$ into the ISM, causing $[\mathrm{C}/\mathrm{O}]$ to rise steeply towards solar values. Furthermore, metal-rich Wolf-Rayet stars lose mass much more efficiently than their low-$Z$ counterparts, resulting in larger C overproduction factors. The contribution of LIMS to the predicted C abundance at the time of Solar System formation ($4.6\,\mathrm{Gyr}$ ago) varies substantially depending on the adopted yield combination: for instance, it reaches about 50 per cent in model MN\,N13, whereas it drops to 20 per cent in model AT\,R150.

Figure~\ref{fig:NO_OH} displays the observational compilation (left panel) and our GCE model tracks (right panel) for $[\mathrm{N}/\mathrm{O}]$ as a function of $[\mathrm{O}/\mathrm{H}]$ in the solar neighbourhood. The empirical distribution displays two main regimes. At low metallicities, $[\mathrm{O}/\mathrm{H}] < -1.7$, the observational data form a broad plateau centred around $[\mathrm{N}/\mathrm{O}] \approx -0.8$, albeit with substantial scatter. The flat behaviour of $[\mathrm{N}/\mathrm{O}]$ at low $[\mathrm{O}/\mathrm{H}]$ is the classic signature of primary nitrogen production. Specifically, the predicted run of $[\mathrm{N}/\mathrm{O}]$ at early times (right-hand panel of Fig.~\ref{fig:NO_OH}) depends critically on the treatment of rotational mixing in massive stars as well as on the assumed mass threshold for direct collapse to black holes. Models relying on non-rotating massive star yields synthesize nitrogen purely as a secondary element (i.e. proportional to the initial CNO abundance), severely underpredicting $[\mathrm{N}/\mathrm{O}]$ at the lowest metallicities. In contrast, models incorporating rotationally induced mixing in massive stars trigger primary $^{14}\mathrm{N}$ synthesis via the CNO cycle by diffusing $^{12}\mathrm{C}$ produced in the He-burning core up to the base of the outer H-burning shell. In particular, the reference \texttt{FRANEC} rotating models (Set~R) naturally reproduce the observed low-metallicity plateau. The rotating \texttt{GENEC} models predict an extreme overproduction of nitrogen at very low metallicities, exceeding $[\mathrm{N}/\mathrm{O}] \approx +0.5$.

The large spread of $[\mathrm{N}/\mathrm{O}]$ ratios at intermediate metallicities ($-1.8 < [\mathrm{O}/\mathrm{H}] < -1.3$) seen in the left panel of Fig.~\ref{fig:NO_OH} is attributable to the presence of several stars with upper/lower limits on their O and N abundances \citep{rossi2024}. At metallicities above $[\mathrm{O}/\mathrm{H}] \approx -1.5$, $[\mathrm{N}/\mathrm{O}]$ begins a gradual rise driven by the delayed contribution of intermediate-mass stars ($4 \le m/M_\odot \le 8$). During the AGB phase, these stars undergo hot bottom burning, which synthesizes primary N with an efficiency that decreases with increasing metallicity. A steeper inflection occurs starting around $[\mathrm{O}/\mathrm{H}] \approx -0.7$, where secondary N production in both low-mass AGB stars and metal-rich massive stars eventually raises the ratio to solar or even super-solar values in high-metallicity disc dwarfs \citep{suarezandres2016}. A few models (AT\,R150, MN\,N13, and the Set~R models) successfully capture the onset and slope of this high-metallicity trend.

    \begin{figure*}
    \centering
    \includegraphics[width=\columnwidth]{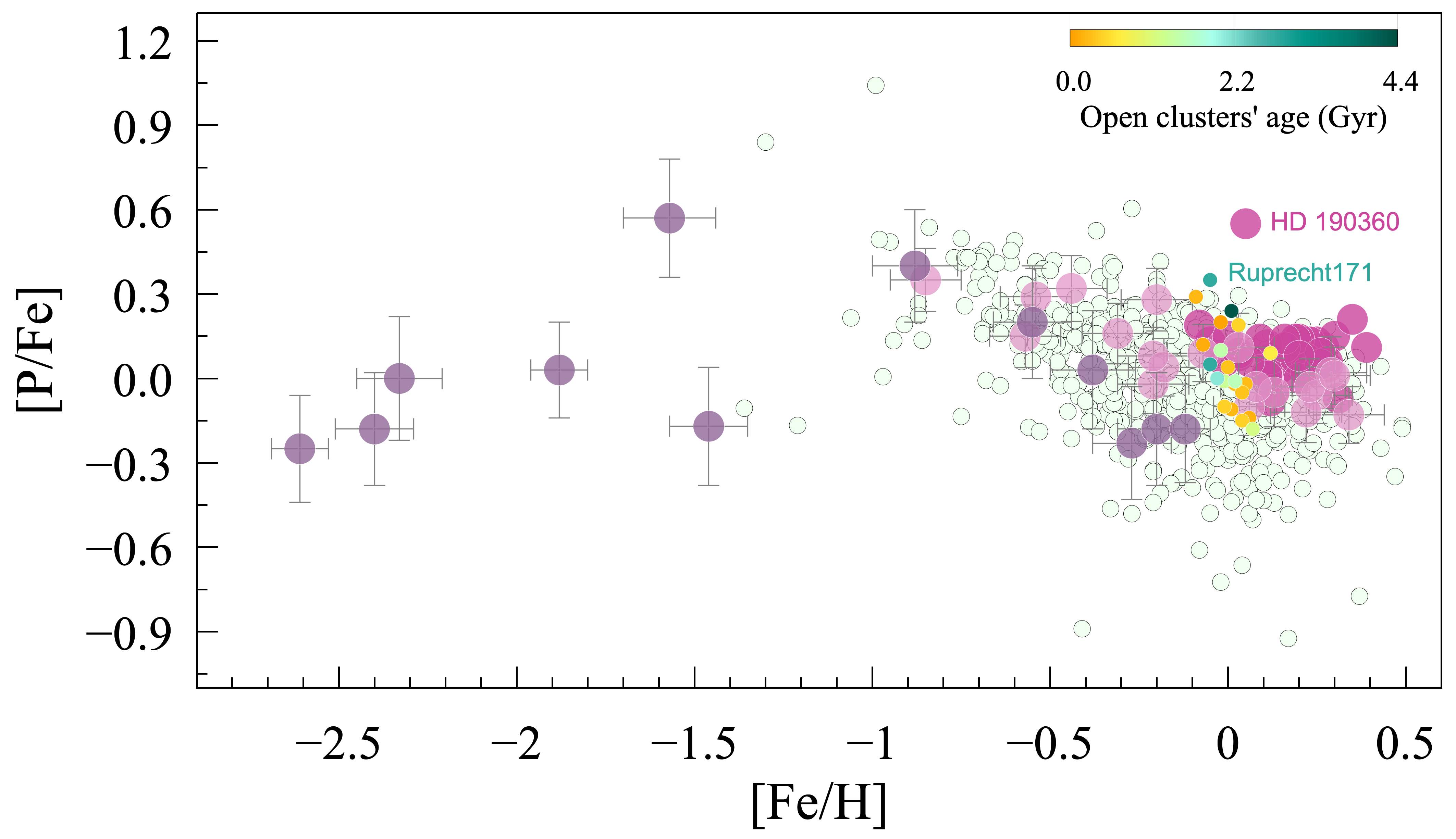}
    \includegraphics[width=\columnwidth]{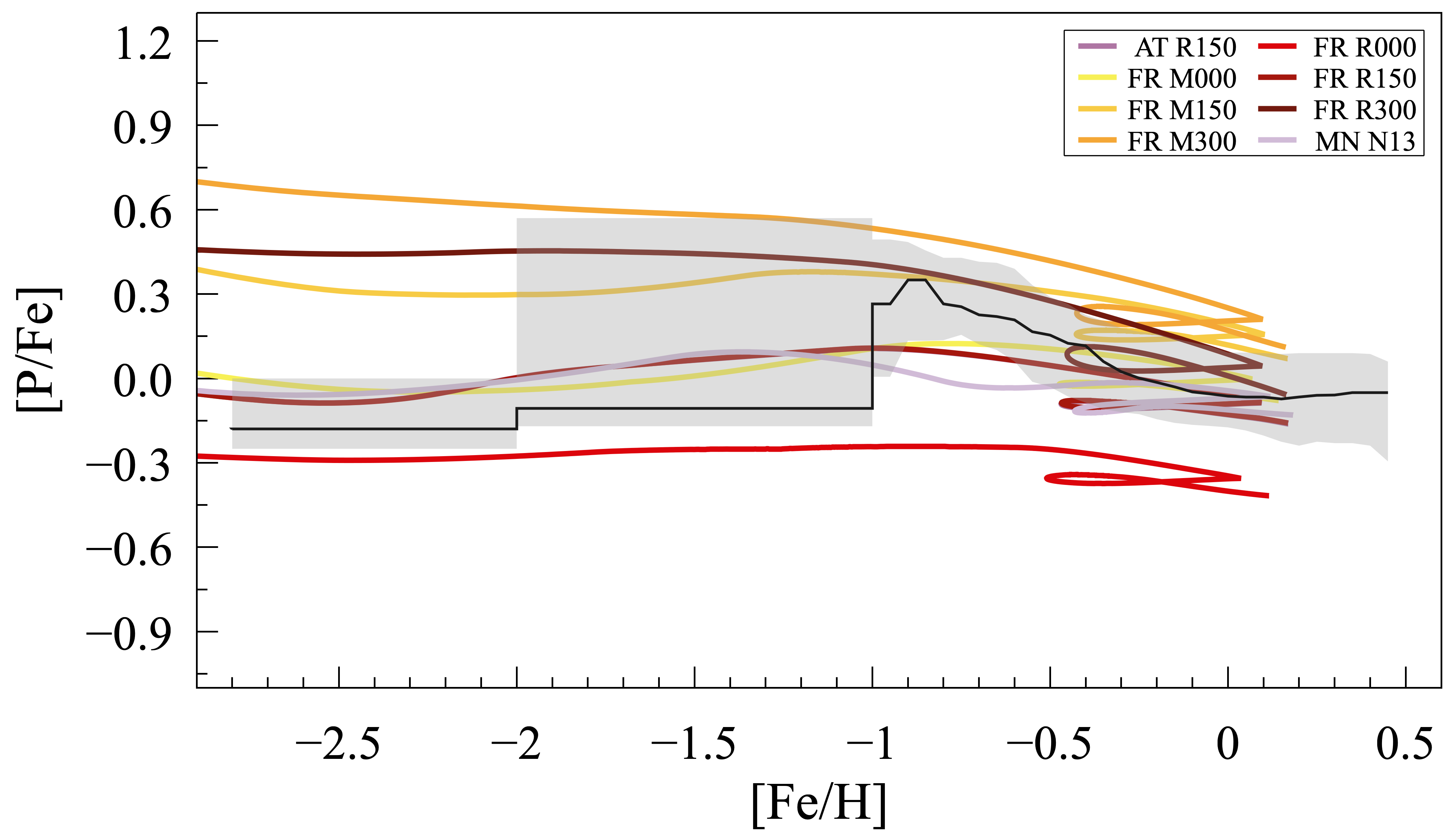}
    \caption{\textit{Left:} $[\mathrm{P}/\mathrm{Fe}]$ vs. $[\mathrm{Fe}/\mathrm{H}]$ for solar-neighbourhood field stars from \citet[][purple circles]{roederer2014}, \citet[][pink circles]{caffau2011}, \citet[][magenta circles]{caffau2019}, and \citet[][pale-green circles]{gakis2026}. Age-coded markers show average abundances of open clusters \citep{jian2026}. \textit{Right:} GCE model predictions (colour-coded as in Fig.~\ref{fig:CO_OH}) and median observational trend computed using a $0.35\,\mathrm{dex}$ sliding window, with broader fixed bins for the sparsely populated $[\mathrm{Fe}/\mathrm{H}] < -1$ regime.}
    \label{fig:PFe_FeH}
    \end{figure*}

    \begin{figure*}
    \centering
    \includegraphics[width=0.49\textwidth]{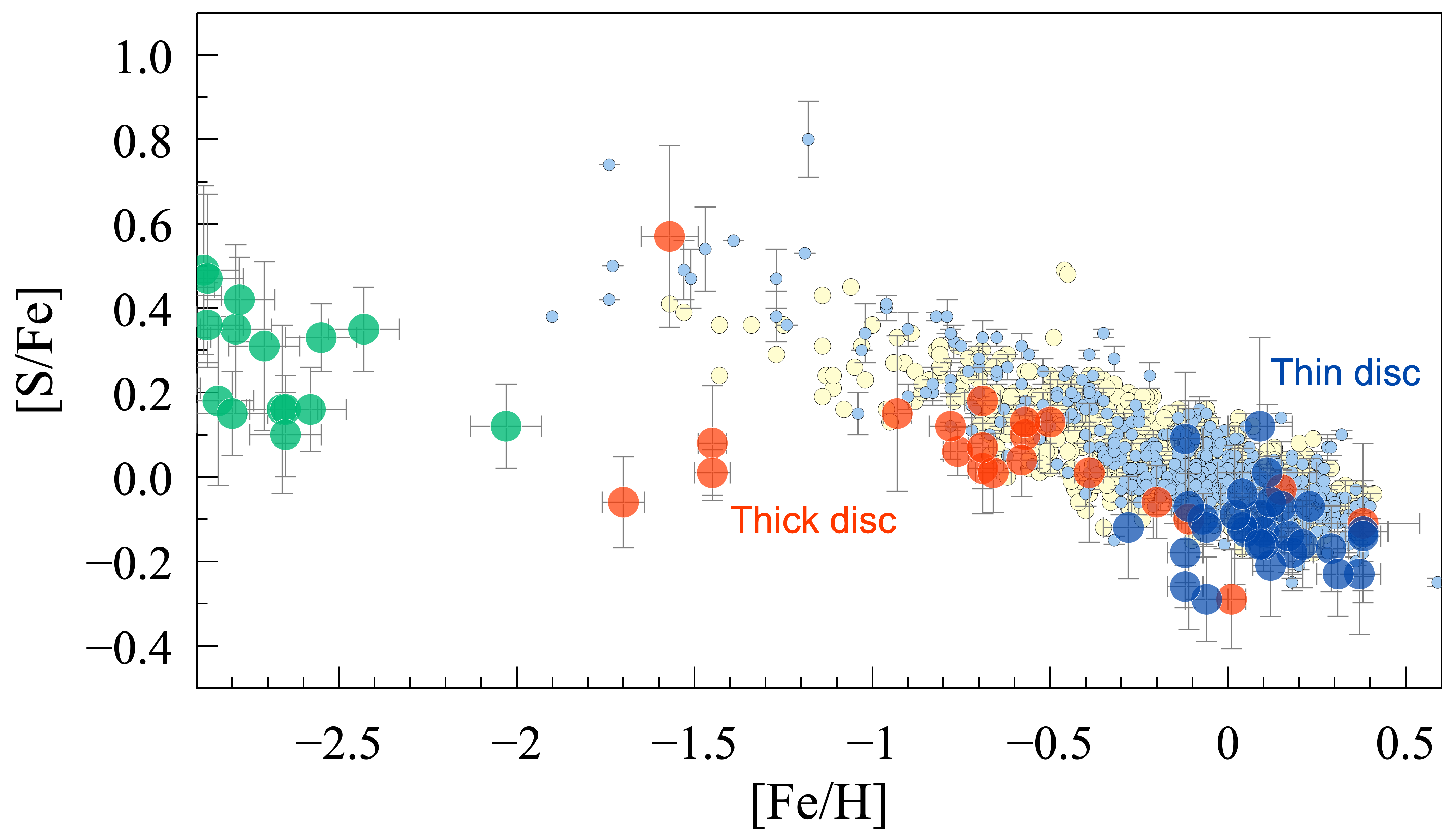}
    \includegraphics[width=0.49\textwidth]{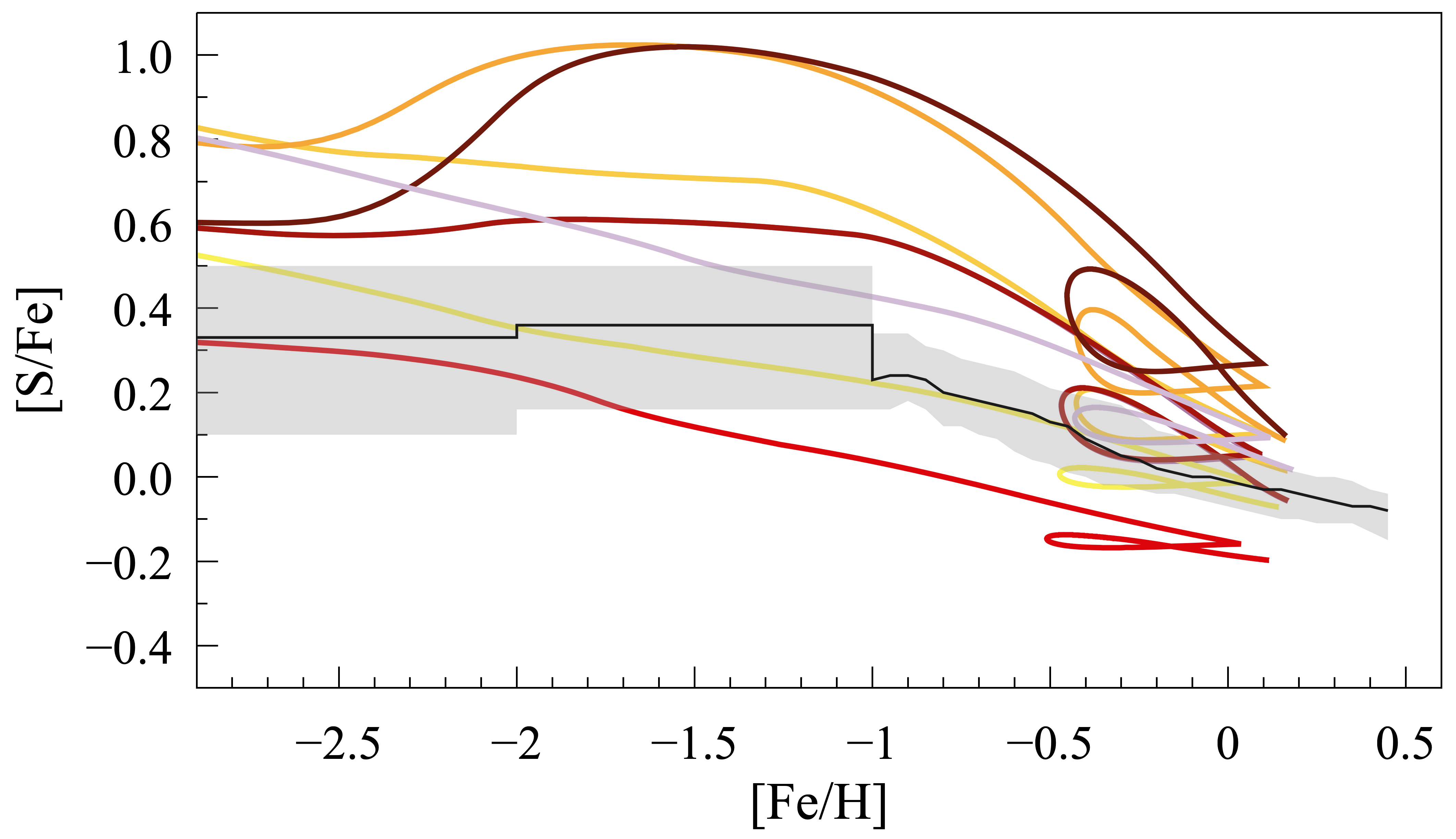}
    \caption{Same as Fig.~\ref{fig:PFe_FeH}, but for $[\mathrm{S}/\mathrm{Fe}]$ vs. $[\mathrm{Fe}/\mathrm{H}]$. Data are from \citet[][green circles]{spite2011}, \citet[][orange and blue circles for thick- and thin-disc F--G stars, respectively]{lucertini2022}, \citet[][small light-blue circles]{perdigon2021}, and \citet[][small cream circles]{bensby2026}.}
    \label{fig:SFe_FeH}
    \end{figure*}

In Fig.~\ref{fig:PFe_FeH}, we present the observational data alongside our GCE model predictions for $[\mathrm{P}/\mathrm{Fe}]$ as a function of $[\mathrm{Fe}/\mathrm{H}]$\footnote{Here and in the remainder of this section, we switch from oxygen to iron as our reference metallicity tracer, since requiring simultaneous oxygen determinations would drastically reduce the available sample size for P, S, and the isotopic ratios.}. The empirical distribution exhibits a plateau at $[\mathrm{P}/\mathrm{Fe}] \approx -0.2$ below $[\mathrm{Fe}/\mathrm{H}] = -2$, followed by a rise and subsequent decline at $[\mathrm{Fe}/\mathrm{H}] > -1$. Non-LTE corrections to P abundances of cool stars over a wide metallicity range do not qualitatively alter the overall P abundance distribution, with the primary effect being a tighter clustering of $[\mathrm{P}/\mathrm{Fe}]$ ratios around $[\mathrm{P}/\mathrm{Fe}] \approx 0$ at low metallicities \citep{andrievsky2026}. In massive stars, $^{31}\mathrm{P}$ is synthesized primarily during explosive Ne burning, mainly from $^{27}\mathrm{Al}$ via the $^{27}\mathrm{Al}(\alpha,\gamma)^{31}\mathrm{P}$ and $^{27}\mathrm{Al}(\alpha,p)^{30}\mathrm{Si}(p,\gamma)^{31}\mathrm{P}$ reaction channels. In turn, $^{27}\mathrm{Al}$ is produced as a primary element through the sequence $^{20}\mathrm{Ne}(\alpha,\gamma)^{24}\mathrm{Mg}(\alpha,p)^{27}\mathrm{Al}$; consequently, $^{31}\mathrm{P}$ exhibits essentially primary behaviour. In low-mass AGB stars ($m \approx 1$--$3\,M_\odot$), $^{31}\mathrm{P}$ is synthesized via slow neutron capture ($s$-process) on $^{30}\mathrm{Si}$, whereas in certain intermediate-to-massive AGB models, additional synthesis channels open up via proton or alpha captures (e.g. on $^{27}\mathrm{Al}$). Since the \texttt{ATON} grid lacks phosphorus, its AGB yield is set to zero in model AT\,R150. Additionally, P evolution predicted with \texttt{GENEC} massive-star yields is not discussed, given the yield uncertainties noted in Sect.~\ref{sec:hmnuc}.

The right-hand panel of Fig.~\ref{fig:PFe_FeH} reveals that GCE model predictions for $[\mathrm{P}/\mathrm{Fe}]$ depend strongly on the assumptions regarding the explodability of massive stars, on the inclusion of rotation, and on the inclusion of the mixing-fallback phenomenon occurring during the explosion. In general, at low metallicities the implementation of both Set~M and Set~R leads to relatively flat $[\mathrm{P}/\mathrm{Fe}]$ trajectories, as is expected for a primary element, before declining towards higher metallicities owing to delayed Fe production from SNe~Ia. However, the R models generally predict a lower $[\mathrm{P}/\mathrm{Fe}]$ ratio compared to the M models, due to the assumption that the most massive stars collapse directly into a black hole. Stellar rotation, on the contrary, tends to increase the $[\mathrm{P}/\mathrm{Fe}]$ ratio. Looking at the specific models, FR\,M000 and the moderately rotating FR\,R150 align remarkably well with the median observational trend within its 16$^\mathrm{th}$--84$^\mathrm{th}$ percentile dispersion envelope across all metallicities. Model FR\,R000 (non-rotating Set~R) predicts a significantly suppressed ratio ($[\mathrm{P}/\mathrm{Fe}] \approx -0.3$ in the early Galaxy and $[\mathrm{P}/\mathrm{Fe}] \approx -0.4$ in the thin disc), while models FR\,M150, FR\,M300, and FR\,R300 lead to a dramatic overproduction of P at low metallicities, overshooting the empirical halo baseline. Overall, the observed $[\mathrm{P}/\mathrm{Fe}]$ behaviour in the solar neighbourhood suggests that rotationally induced yield enhancements for odd-$Z$ elements at low metallicity must be substantially milder than assumed in high-rotation grids, with model FR\,R150 being the sole exception. Within the framework of our GCE models, the contribution of LIMS to the predicted P abundance at the time of Solar System formation ($4.6\,\mathrm{Gyr}$ ago) is negligible, amounting to 2 per cent in model MN\,N13 and $< 1$ per cent in model FR\,R150.

Sulphur is a classic $\alpha$-element primarily synthesized in massive stars, where it is produced in roughly equal amounts during explosive oxygen and incomplete explosive silicon burning, with a minor contribution from explosive neon-carbon burning. At later times, an additional fraction comes from SNe~Ia, which account for between one-quarter and one-third of the predicted S abundance at the time of Solar System formation in our models. In Fig.~\ref{fig:SFe_FeH}, we present the observational compilation alongside our GCE model trajectories for $[\mathrm{S}/\mathrm{Fe}]$ as a function of $[\mathrm{Fe}/\mathrm{H}]$. The empirical distribution displays the canonical $\alpha$-element trend: at low metallicities, the observational data define a plateau, while for $[\mathrm{Fe}/\mathrm{H}] > -1$, delayed Fe enrichment from SNe~Ia creates a sharp inflection point, causing $[\mathrm{S}/\mathrm{Fe}]$ to decline steadily towards solar values. GCE model predictions vary markedly depending on the adopted massive-star yields (Fig.~\ref{fig:SFe_FeH}, right panel): model FR\,M000 successfully reproduces the observed behaviour within its dispersion envelope, while models including fast rotators predict an overproduction of sulphur at intermediate metallicities, reaching values up to $[\mathrm{S}/\mathrm{Fe}] \approx +1$ (assuming complete black-hole collapse for stars more massive than $25\,M_\odot$ mitigates S overproduction by locking the regions where S is produced inside the remnant). In these models, rotation-induced mixing leads to the formation of a larger CO core, making the innermost regions of the star significantly more compact. As a consequence, a greater amount of mass is exposed to the thermodynamic conditions required for explosive nucleosynthesis, thereby drastically enhancing the synthesis of S (and, more generally, of all the elements and isotopes primarily produced during the explosion). Moreover, the lower-mass rotating models (i.e. the $13$ and $15\,M_\odot$ models) at low metallicities undergo C--O shell merging in the very late stages prior to the explosion. This phenomenon substantially enhances the production of all the elements synthesized during O-shell burning (including S), as discussed in detail by \citet{Roberti+2025}. Because sulphur is not synthesized in LIMS, the observed $[\mathrm{S}/\mathrm{Fe}]$ track provides an unpolluted baseline for massive-star nucleosynthesis, favouring non-rotating or moderately rotating CCSN yields.

    \begin{figure*}
    \centering
    \includegraphics[width=0.49\textwidth]{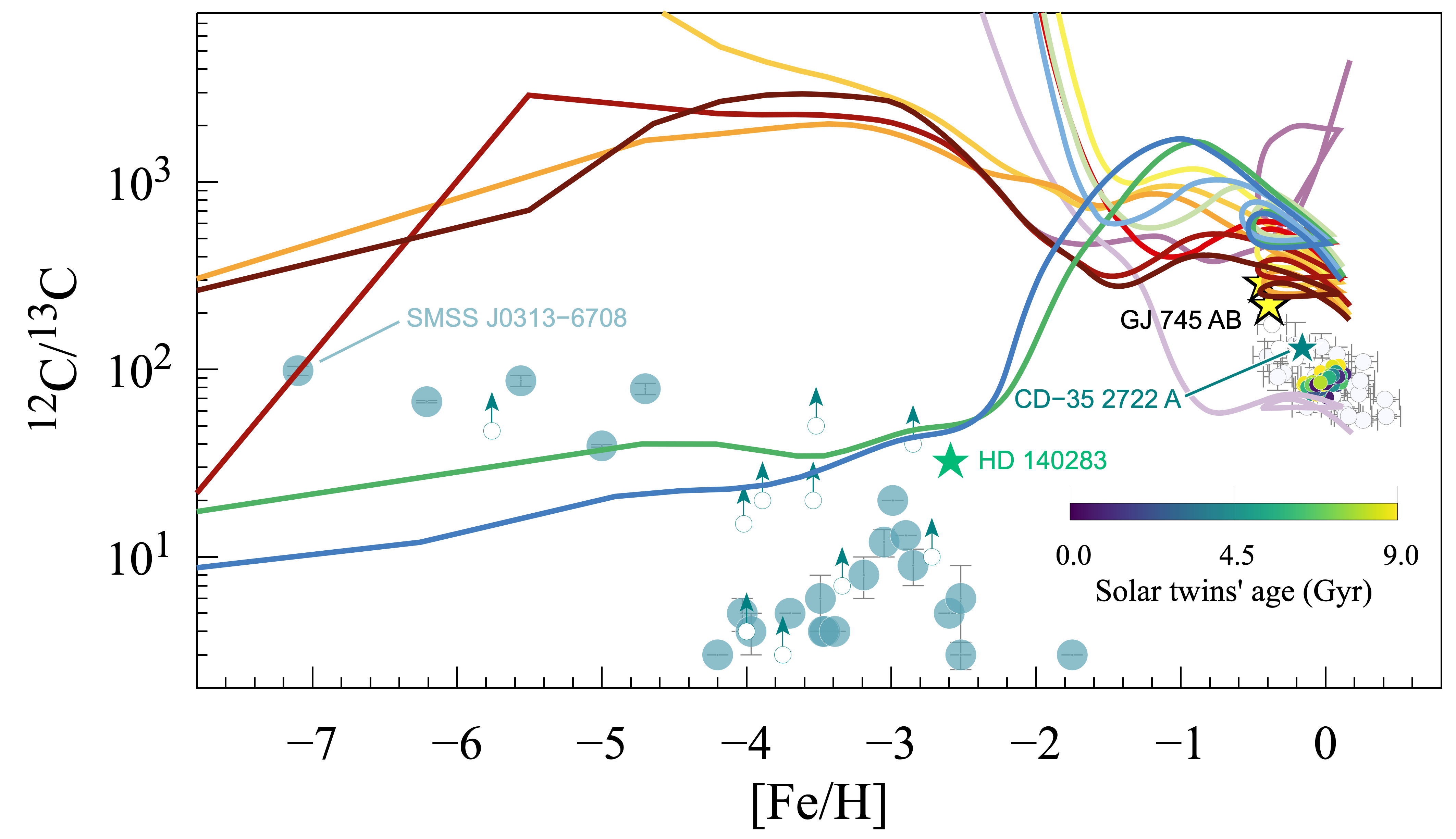}
    \includegraphics[width=0.49\textwidth]{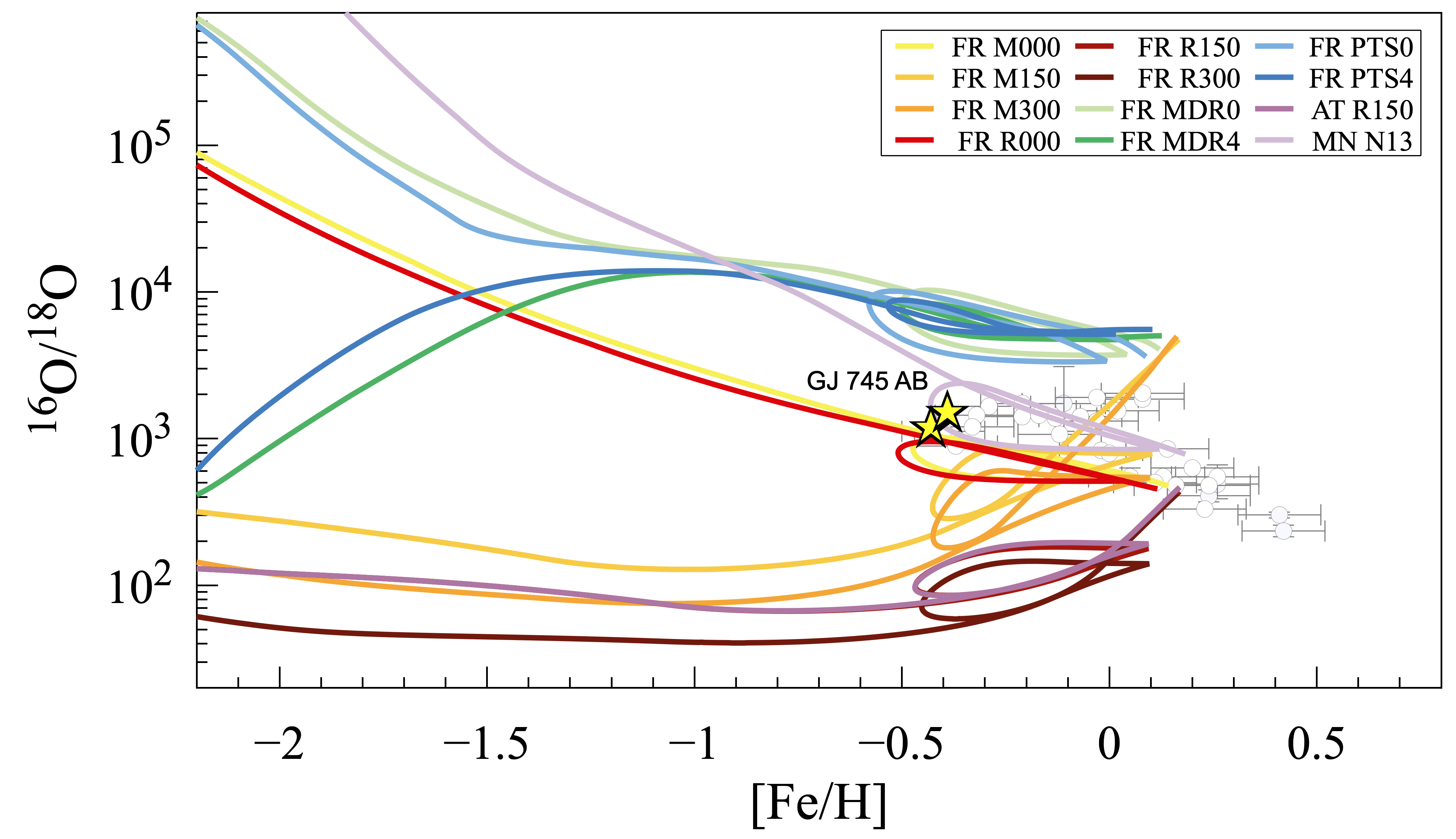}
    \caption{ Carbon \textit{(left)} and oxygen \textit{(right)} isotopic ratios vs. metallicity in the solar neighbourhood. Coloured solid lines show GCE model predictions. Data include age-coded solar twins \citep{botelho2020}, M dwarfs \citep[][yellow stars, white circles, and teal star, respectively]{crossfield2019,gonzalezpicos2025,wang2026}, CEMP-no stars \citep[][filled teal circles for measurements and open circles with arrows for lower limits]{molaro2023}, and the ancient turnoff star HD\,140283 \citep[][green star]{spite2021}.} 
    \label{fig:COiso_FeH}
    \end{figure*}

    \begin{figure}
    \centering
    \includegraphics[width=0.49\textwidth]{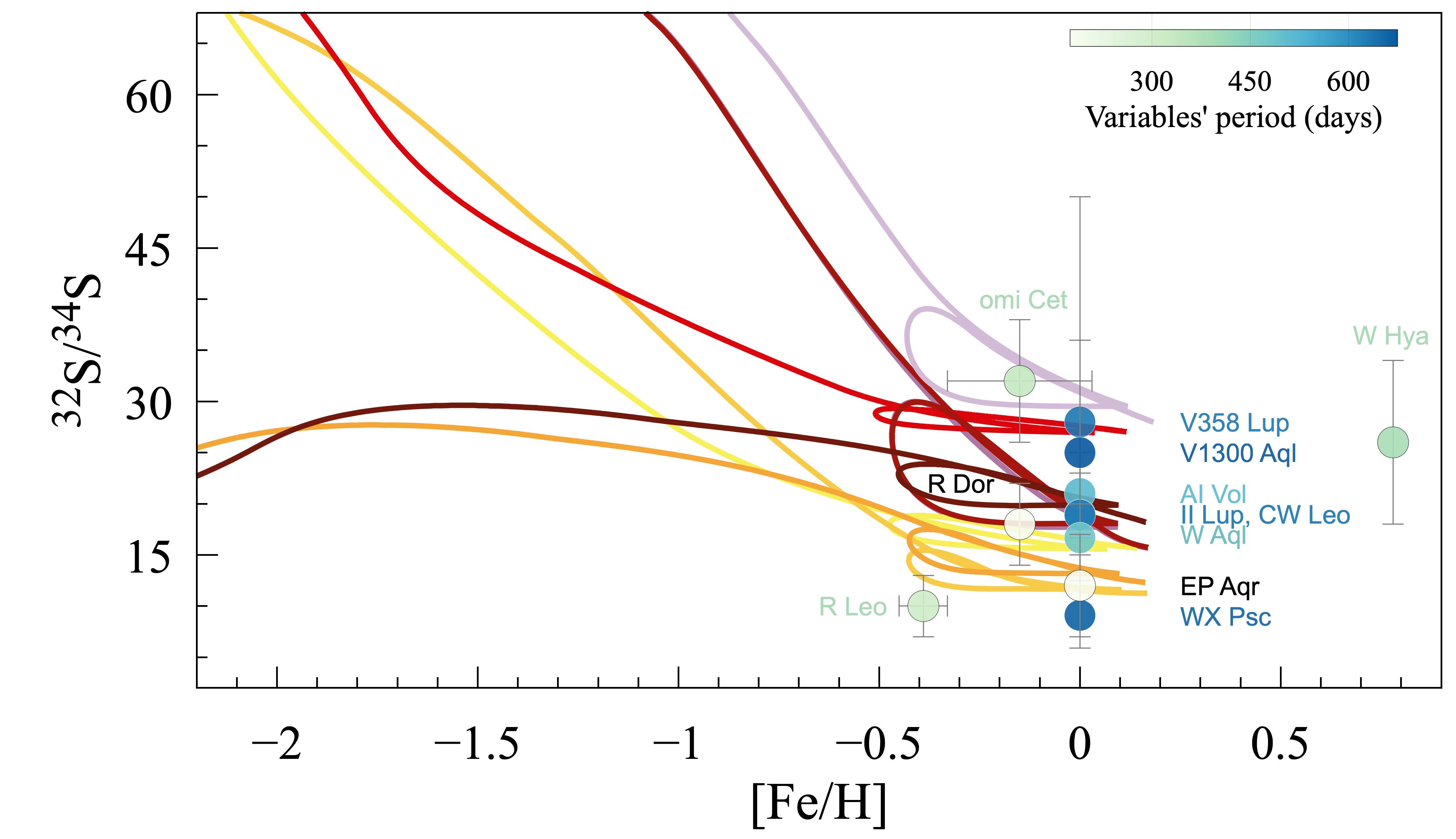}
    \caption{$^{32}\mathrm{S}/^{34}\mathrm{S}$ vs. $[\mathrm{Fe}/\mathrm{H}]$ in the solar neighbourhood. Circumstellar envelope measurements of local AGB stars \citep[filled circles,][]{unnikrishnan2024,wallstroem2024,gorai2025} are colour-coded by pulsation period (longer periods trace younger ages). Stars plotted at $[\mathrm{Fe}/\mathrm{H}] = 0$ lack precise metallicities and are assigned nominal solar values. Solid lines display GCE model predictions.}
    \label{fig:Siso_FeH}
    \end{figure}

Figure~\ref{fig:COiso_FeH} compares our theoretical predictions with C and O isotopic ratios measured in nearby unevolved stars. GCE model predictions vary dramatically in the halo regime depending on massive-star physics. Models adopting non-rotating massive-star yields, where $^{13}\mathrm{C}$ is synthesized purely as a secondary element, predict C isotopic ratios exceeding $10^3$ at early times, in sharp contrast with observations (Fig.~\ref{fig:COiso_FeH}, left panel). However, halo stars with high $^{12}\mathrm{C}/^{13}\mathrm{C}$ ratios, if present, would display weaker $^{13}\mathrm{C}$ features that are harder to detect; thus, an observational selection bias cannot be ruled out. Efficient rotationally induced mixing at low metallicities transports core-synthesized $^{12}\mathrm{C}$ into the H-burning shell, triggering primary $^{13}\mathrm{C}$ synthesis. This lowers the predicted early $^{12}\mathrm{C}/^{13}\mathrm{C}$ ratio by orders of magnitude in models implementing rotating stars. For $[\mathrm{Fe}/\mathrm{H}] > -2$, the run of $^{12}\mathrm{C}/^{13}\mathrm{C}$ is governed by a complex interplay among multiple processes: while any contribution to primary $^{13}\mathrm{C}$ from massive stars is fading, $^{12}\mathrm{C}$ injection from high- and low-mass stars, alongside $^{13}\mathrm{C}$ enhancement driven by hot bottom burning in intermediate-mass stars and secondary production in Wolf-Rayet and low-mass stars, take over. The net balance of these physical processes leads to different model trajectories, depending on the adopted yield prescriptions: model AT\,R150 predicts a continuous rise in $^{12}\mathrm{C}/^{13}\mathrm{C}$ throughout most of the disc evolution, model MN\,N13 yields $^{12}\mathrm{C}/^{13}\mathrm{C}$ ratios compatible with the lowest values observed in local stars, and all other models predict ratios in excess of the empirical baseline, hinting at the need for another source of $^{13}\mathrm{C}$ at late times (see Sect.~\ref{sec:mix}).

The right-hand panel of Fig.~\ref{fig:COiso_FeH} illustrates the run of $^{16}\mathrm{O}/^{18}\mathrm{O}$ against $[\mathrm{Fe}/\mathrm{H}]$ in the solar neighbourhood. While $^{16}\mathrm{O}$ is a primary element produced predominantly during He burning in massive stars, $^{18}\mathrm{O}$ is synthesized via $\alpha$-captures on $^{14}\mathrm{N}$ during core and shell He burning---although easily destroyed in LIMS, it can survive in massive stars \citep{hirschi2007,limongi2018}. Fast rotation, in particular, enhances primary $^{14}\mathrm{N}$ production in low-metallicity massive stars, which subsequently feeds primary $^{18}\mathrm{O}$ synthesis during He burning. This significantly depresses the early $^{16}\mathrm{O}/^{18}\mathrm{O}$ ratio, down to $\sim 10^2$ in some models. Near solar metallicity, secondary $^{18}\mathrm{O}$ production dominates and most models converge within $^{16}\mathrm{O}/^{18}\mathrm{O} \approx 400$--$1500$, comfortably enveloping the M-dwarf measurements. Extending $^{16}\mathrm{O}/^{18}\mathrm{O}$ determinations to M-dwarf stars with $[\mathrm{Fe}/\mathrm{H}] < -0.5$ would provide crucial constraints on stellar and GCE models.

    \begin{figure*}
    \centering
    \includegraphics[width=0.49\textwidth]{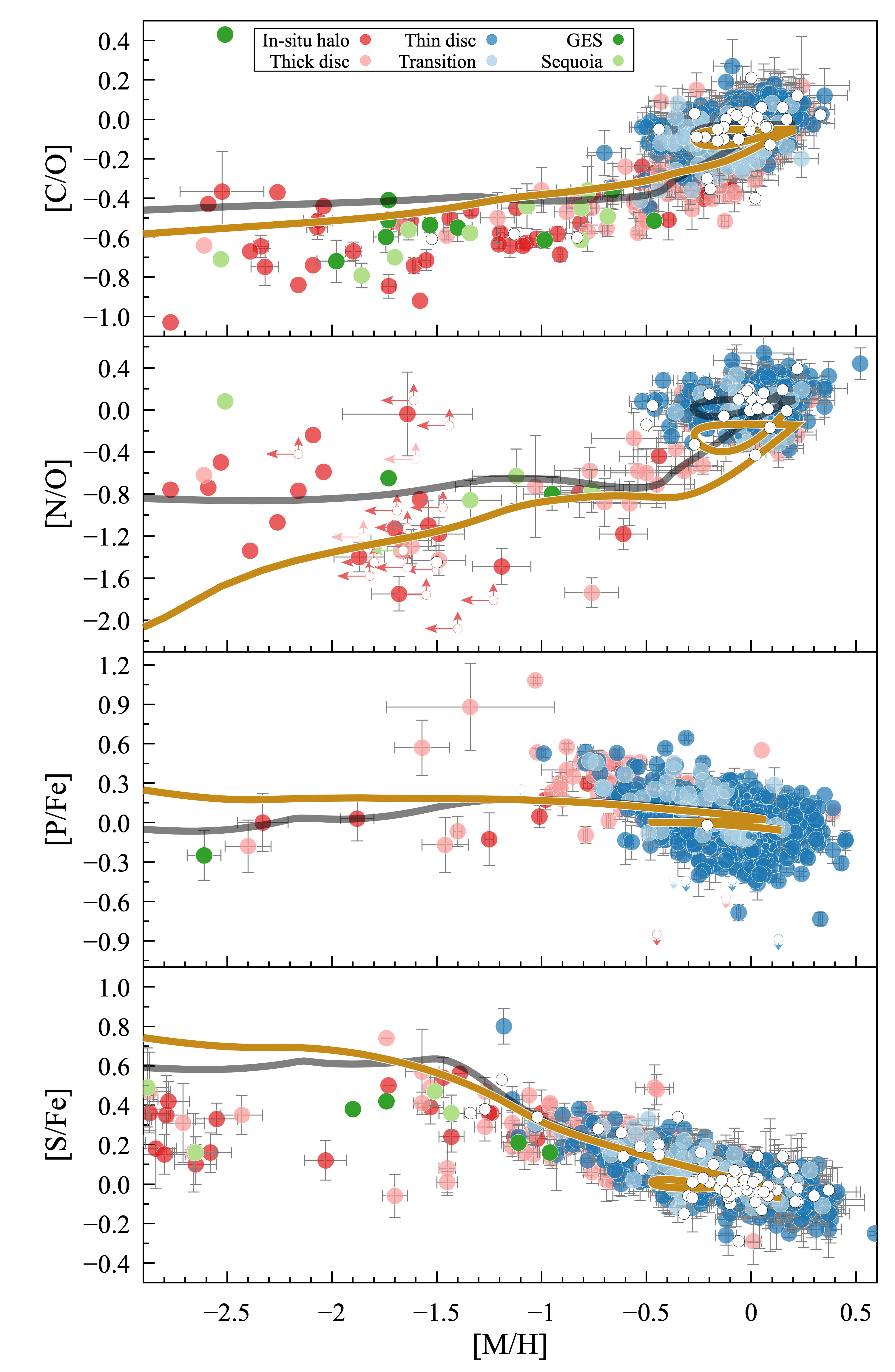}
    \includegraphics[width=0.49\textwidth]{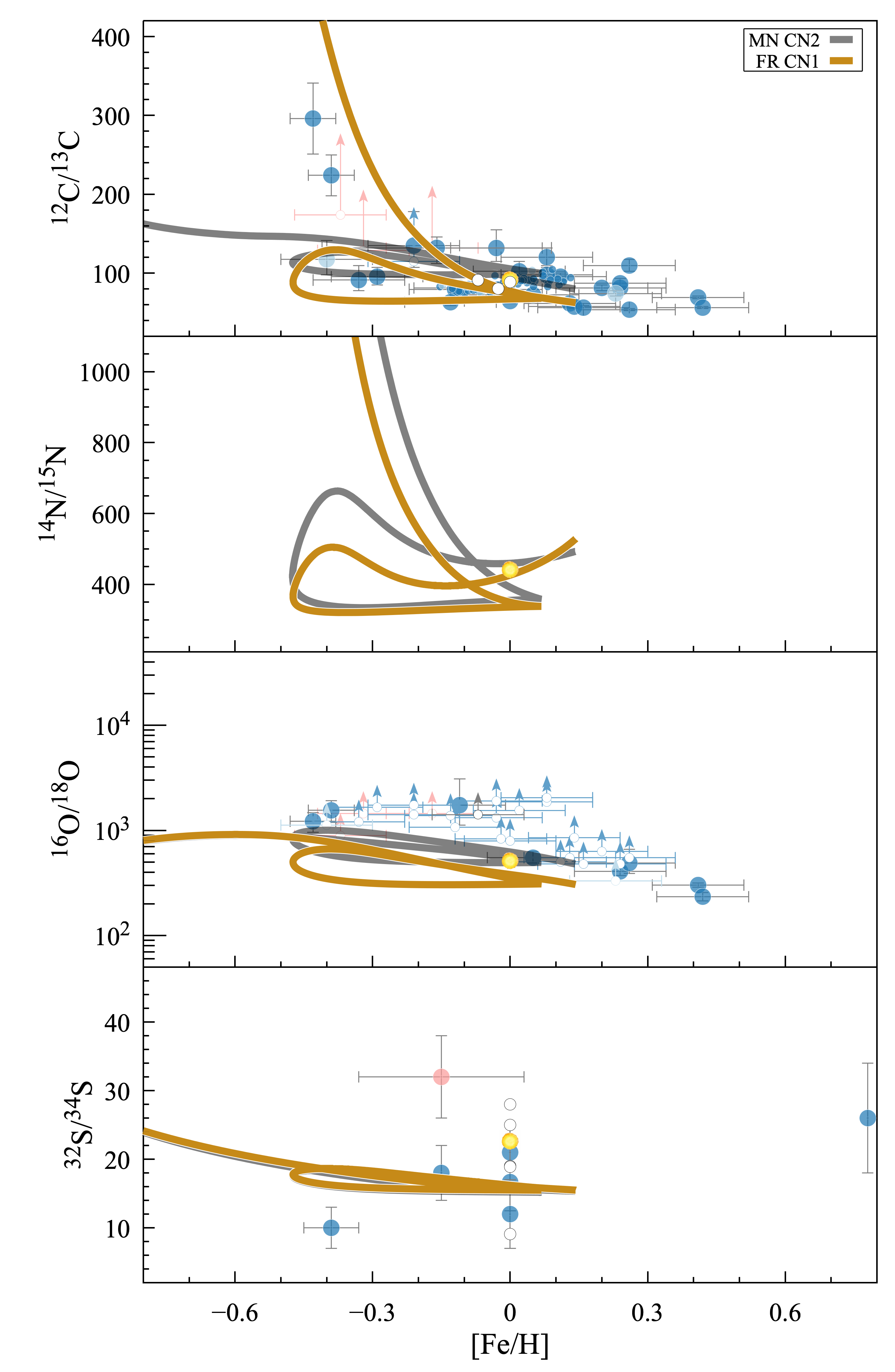}
    \caption{Evolution of CNOPS elemental \textit{(left)} and isotopic \textit{(right)} ratios in the solar neighbourhood predicted by models FR\,CN1 and MN\,CN2 (solid lines). Model predictions are compared with our data compilation (see Sect.~\ref{sec:data_SN}). The stars are homogeneously classified into in-situ halo, thick disc, thin disc, transition, and accreted components (see legend in the top-left panel and Appendix~\ref{app:kin}); white circles denote kinematically unclassified stars, and open circles with arrows indicate upper or lower limits. The yellow circles in the right panels mark the (proto)solar ratios.}
    \label{fig:novae}
    \end{figure*}

In Fig.~\ref{fig:Siso_FeH}, we present the predicted local evolution of $^{32}\mathrm{S}/^{34}\mathrm{S}$ alongside measurements in the circumstellar envelopes of a few local AGB stars. Because sulphur isotopes remain unaltered by internal processes in LIMS, AGB envelope ratios directly record the ISM composition at the time of progenitor birth. Observed ratios in local AGB envelopes cluster between $^{32}\mathrm{S}/^{34}\mathrm{S} \approx 10$ and $35$ \citep[the value of the ratio in the protosolar nebula is $22.6$,][]{ding1999}. The dominant isotope, $^{32}\mathrm{S}$, is a primary species formed via incomplete explosive silicon burning and explosive oxygen burning in CCSNe and SNe~Ia. Conversely, $^{34}\mathrm{S}$ has both a primary component from explosive oxygen burning and a secondary component (from $^{33}\mathrm{S}(n,\gamma)^{34}\mathrm{S}$) from explosive Ne burning that becomes increasingly important at higher metallicities. Most GCE models predict high initial ratios ($^{32}\mathrm{S}/^{34}\mathrm{S} > 60$) that decline steadily as secondary $^{34}\mathrm{S}$ production rises with metallicity. Models incorporating yields from the fastest massive rotators (FR\,M300, FR\,R300) are notable exceptions, producing flatter trajectories across their entire evolution, partly due to the contribution from stars that undergo C--O shell merging \citep{Roberti+2025}. During the thin-disc phase, all GCE tracks flatten and converge into the narrow range ($^{32}\mathrm{S}/^{34}\mathrm{S} \approx 10$--$30$) defined by AGB circumstellar envelopes. At the time of Solar System formation, SNe~Ia contribute approximately 25 per cent of $^{32}\mathrm{S}$ and 15 per cent of $^{34}\mathrm{S}$ in model MN\,N13, compared to nearly 20 per cent and 5 per cent in model FR\,M300.

\subsection{Baseline models}
\label{sec:mix}

Low-metallicity, massive fast rotators are essential for explaining elevated $[\mathrm{N}/\mathrm{O}]$ ratios in the halo \citep{meynet2002,chiappini2006,romano2010,romano2019,prantzos2018}. While boosting primary $^{14}\mathrm{N}$, these stars also enhance $^{13}\mathrm{C}$ production relative to non-rotating models. Yet, GCE models tuned to the $[\mathrm{N}/\mathrm{O}]$ plateau tend to overestimate $^{12}\mathrm{C}/^{13}\mathrm{C}$ ratios in metal-poor stars (see previous section). This discrepancy may stem from mild $^{12}\mathrm{C}$ overproduction in low-metallicity massive star models (Fig.~\ref{fig:CO_OH}) combined with an observational selection bias against $^{13}\mathrm{C}$-poor stars at $[\mathrm{Fe}/\mathrm{H}] < -2$: indeed, metal-poor stars with $^{12}\mathrm{C}/^{13}\mathrm{C} > 90$ do exist, providing a robust empirical benchmark independent of uncertainties in stellar atmosphere modelling \citep{caffau2025}.

A holistic assessment across the full CNOPS ensemble shows that model FR\,R150 provides a satisfactory fit to all abundance and isotopic ratios for $[\mathrm{Fe}/\mathrm{H}] < -1.5$, beyond which its single-velocity tracks diverge from observations. Assuming a uniform rotational velocity across all masses and metallicities is an unphysical simplification; realistic models clearly require a distribution of initial rotational velocities. However, such a distribution remains poorly constrained observationally \citep[e.g.][]{ramirezagudelo2013,holgado2022} and involves a high degree of freedom in its practical GCE implementation \citep[e.g.][]{prantzos2018,romano2019}.

Guided by empirical evidence for higher rotation velocities in low-metallicity stars and theoretical expectations that direct black hole collapse is facilitated at low $Z$ (owing to weaker winds and larger CO cores), we formulate two composite models:
\begin{itemize}
\item FR\,CN1 (dark ochre lines in Fig.~\ref{fig:novae}): adopts a 50:50 mixture of \citet{limongi2018} Sets~R and M with $v_{\mathrm{ini}} = 150~\mathrm{km\,s^{-1}}$ for $[\mathrm{Fe}/\mathrm{H}] \le -2$, Set~M with $v_{\mathrm{ini}} = 0~\mathrm{km\,s^{-1}}$ for $[\mathrm{Fe}/\mathrm{H}] = -1$ and $0$, and non-rotating yields from \citet{roberti2024} at super-solar metallicities, coupled with FRUITY LIMS yields \citep{cristallo2011,cristallo2015}.
\item MN\,CN2 (grey lines in Fig.~\ref{fig:novae}): implements Set~R with $v_{\mathrm{ini}} = 150~\mathrm{km\,s^{-1}}$ for $[\mathrm{Fe}/\mathrm{H}] \le -3$, a 70:30 mixture of Sets~R and M with $v_{\mathrm{ini}} = 150~\mathrm{km\,s^{-1}}$ at $[\mathrm{Fe}/\mathrm{H}] = -2$, Set~M with $v_{\mathrm{ini}} = 150~\mathrm{km\,s^{-1}}$ at $[\mathrm{Fe}/\mathrm{H}] = -1.5$, and Set~M with $v_{\mathrm{ini}} = 0~\mathrm{km\,s^{-1}}$ for $[\mathrm{Fe}/\mathrm{H}] \ge -1$, combined with Monash LIMS yields \citep{lugaro2012,fishlock2014,karakas2016,karakas2018}.
\end{itemize}
Both models incorporate SN~Ia yields from \citet{iwamoto1999} and nova yields based on the ONe4 model of \citet{jose1998}, empirically adjusted for key isotopes: in model FR\,CN1, the original $^{15}\mathrm{N}$ yield is reduced by a factor of 15; in model MN\,CN2, the original $^{13}\mathrm{C}$ and $^{15}\mathrm{N}$ yields are reduced by one order of magnitude, and the $^{18}\mathrm{O}$ yield is set to zero.

While the choice of LIMS yields does not affect P and S evolution (as expected), Monash yields produce $[\mathrm{C}/\mathrm{O}]$ and $[\mathrm{N}/\mathrm{O}]$ ratios in better agreement with disc observations. Both models FR\,CN1 and MN\,CN2 predict a monotonically decreasing $^{12}\mathrm{C}/^{13}\mathrm{C}$ trend in the disc, aligning with M-dwarf measurements \citep{gonzalezpicos2025} but at variance with solar-twin results \citep{botelho2020}. Crucially, our models reproduce the observed $[\mathrm{P}/\mathrm{Fe}]$ trajectory across almost all metallicities, resolving the P underproduction found in earlier models. In particular, black hole collapse in metal-poor stars above $25~\mathrm{M}_\odot$ ensures near-solar $[\mathrm{P}/\mathrm{Fe}]$ ratios in the halo. Finally, classical novae supply the necessary late-time injection of $^{13}\mathrm{C}$ and $^{15}\mathrm{N}$, bringing C and N isotopic ratios into alignment with local determinations. A few discrepancies remain: the theoretical $[\mathrm{C}/\mathrm{O}]$ and $[\mathrm{S}/\mathrm{Fe}]$ tracks run along the upper envelope of the halo scatter, while the $^{16}\mathrm{O}/^{18}\mathrm{O}$ curves sit near the lower boundary of the M-dwarf data. Furthermore, although the models reproduce the overall $[\mathrm{P}/\mathrm{Fe}]$ trend within the observational uncertainties, they do not capture the mild rise around $[\mathrm{Fe}/\mathrm{H}] \approx -1$ that precedes the decline driven by bulk Fe injection from SNe~Ia. Nevertheless, optimising against the complete CNOPS dataset, rather than single diagnostics, establishes a self-consistent baseline for future work.

A critical next step in testing these baseline models is to confront theoretical predictions with stellar ages. Because metallicity does not map uniquely onto time across distinct Galactic components, evolutionary signatures that overlap or remain hidden in classical $[\mathrm{El}/\mathrm{M}]$ versus $[\mathrm{M}/\mathrm{H}]$ diagrams can be disentangled once reliable ages are available. Indeed, \citet{bensby2026} recently demonstrated that combining homogeneous non-LTE abundances of C, N, O, and S with revised isochrone ages sharply separates old and young disc populations, ranking C, O, and S among the most age-sensitive chemical diagnostics. In Appendix~\ref{app:ages}, we present a preliminary comparison of our FR\,CN1 and MN\,CN2 tracks against the combined literature sample as a function of age; this issue will be investigated in greater depth in forthcoming papers of the SPONGE series, exploiting homogeneous abundances and high-precision evolutionary ages derived for subgiant stars with \textit{Gaia} parallax-based surface gravities.

\subsection{Radial abundance profiles}

    \begin{figure*}
    \centering
    \includegraphics[width=0.49\textwidth]{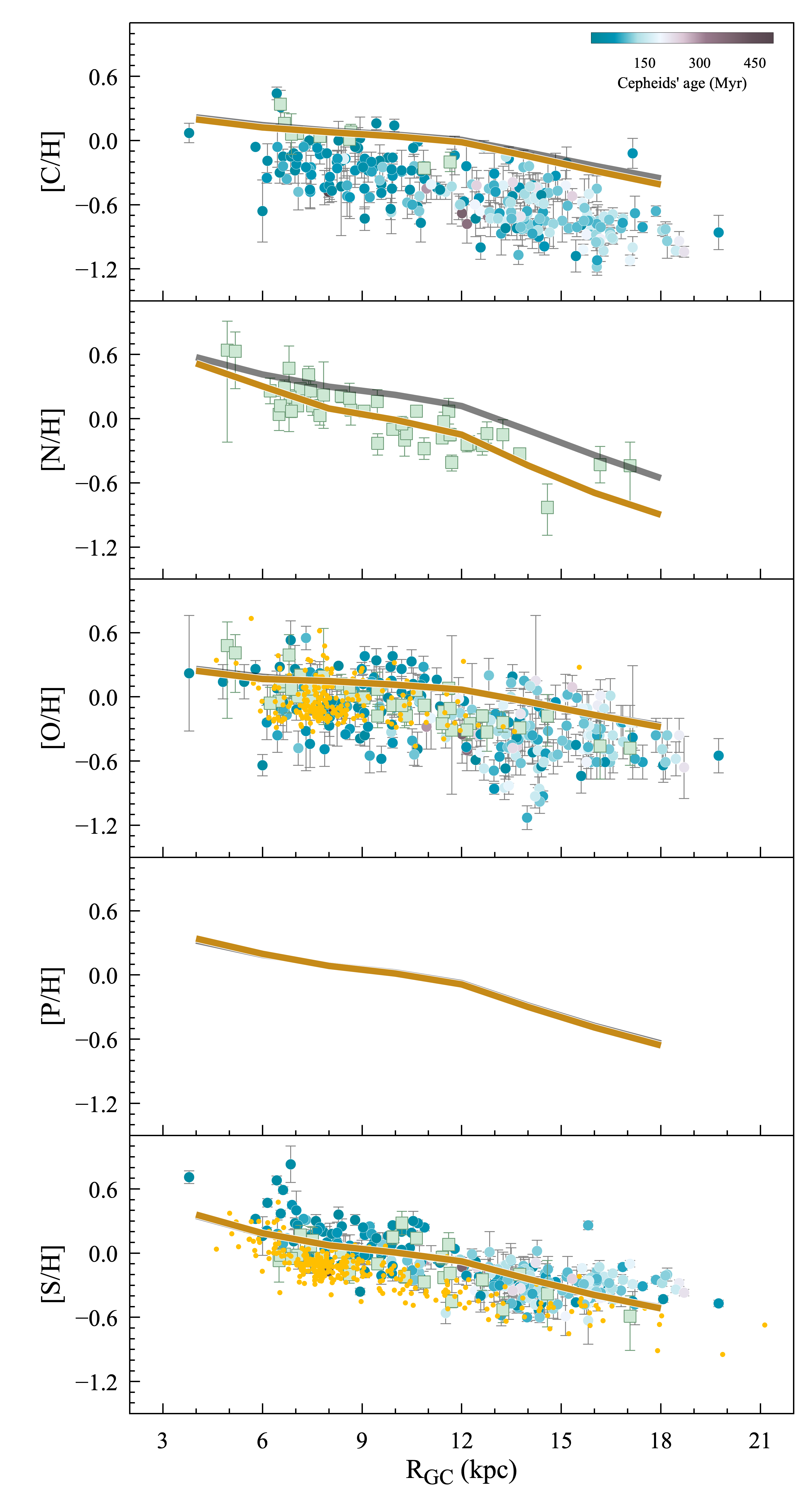}
    \includegraphics[width=0.49\textwidth]{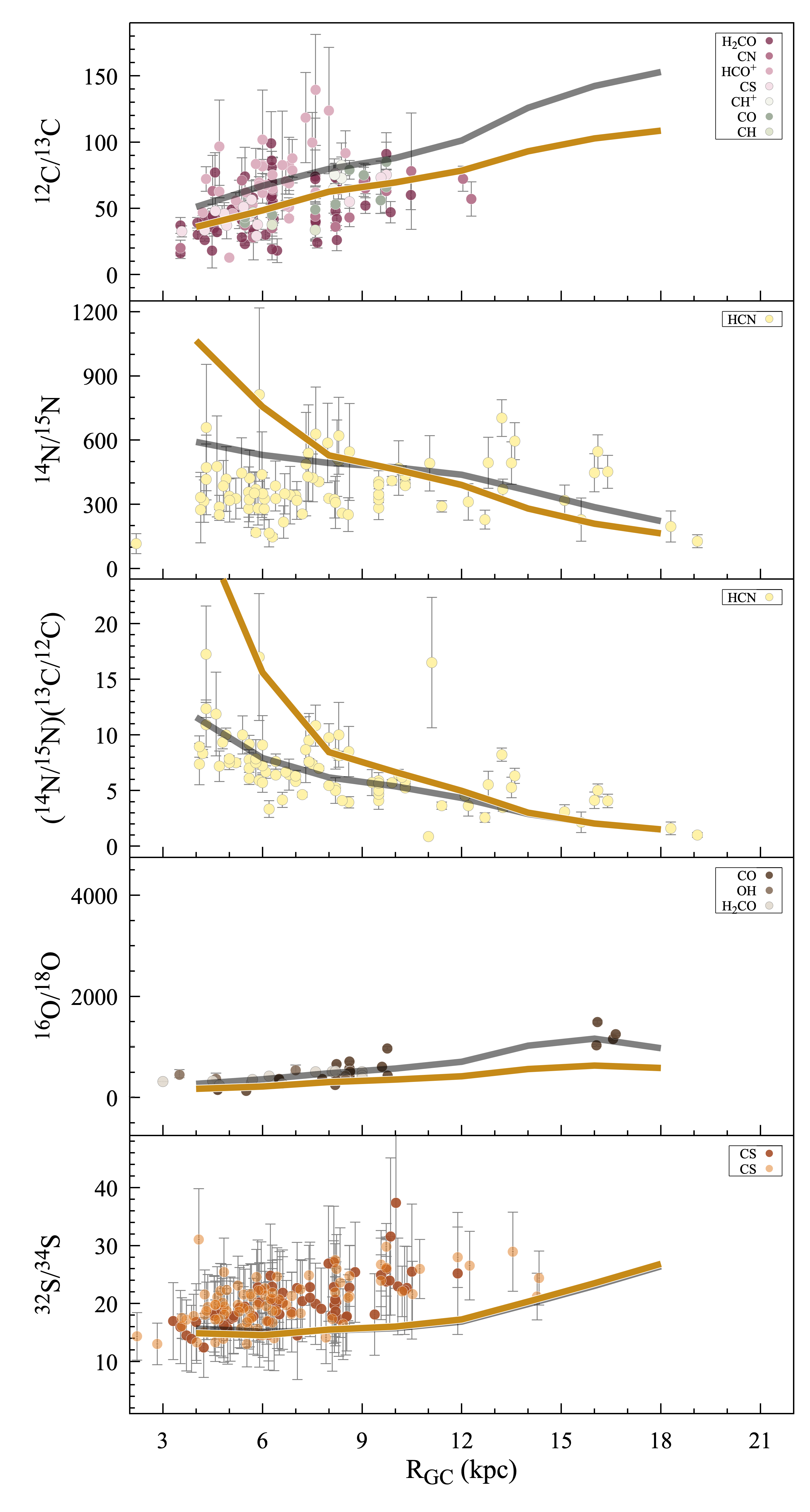}
    \caption{Radial gradients of CNOPS elemental abundances \textit{(left panels)} and isotopic ratios \textit{(right panels)} across the Milky Way disc, compared with predictions from models FR\,CN1 (dark ochre lines) and MN\,CN2 (grey lines). The left panels display LTE \citep[age-coded circles,][]{trentin2023,trentin2024} and non-LTE abundances of classical Cepheids \citep[orange dots,][]{nunnari2026} alongside \ion{H}{ii} region data \citep[green squares,][]{mendezdelgado2022}. The right panels show molecular gas measurements (circles colour-coded by tracer; see Sect.~\ref{sec:data_DISC} for references).}
    \label{fig:grad}
    \end{figure*}

In Fig.~\ref{fig:grad}, we present the spatial distribution of CNOPS elemental abundances (left panels) and key isotopic ratios (right panels) across the Milky Way disc ($R_{\mathrm{GC}} = 4$--$18~\mathrm{kpc}$) predicted by our reference models, FR\,CN1 and MN\,CN2. On the left, model predictions are evaluated against young disc tracers (classical Cepheids and \ion{H}{ii} regions). Overall, the baseline models reproduce the characteristic negative radial abundance gradients of $[\mathrm{C}/\mathrm{H}]$, $[\mathrm{N}/\mathrm{H}]$, $[\mathrm{O}/\mathrm{H}]$, and $[\mathrm{S}/\mathrm{H}]$ (and predict a similar negative slope for $[\mathrm{P}/\mathrm{H}]$, for which radial gradient measurements are not yet available), reflecting the inside-out assembly of the disc. While the models match the observed $[\mathrm{O}/\mathrm{H}]$ and $[\mathrm{S}/\mathrm{H}]$ gradients closely and bracket the $[\mathrm{N}/\mathrm{H}]$ ratios measured in \ion{H}{ii} regions, the predicted $[\mathrm{C}/\mathrm{H}]$ profile sits above the bulk of the Cepheid data.

For the isotopic ratios (right panels of Fig.~\ref{fig:grad}), the predicted radial profiles reflect the primary, secondary, or mixed nature of the underlying isotopes and the spatially varying contributions of LIMS, massive stars, and novae. Whereas $^{12}\mathrm{C}/^{13}\mathrm{C}$, $^{16}\mathrm{O}/^{18}\mathrm{O}$, and $^{32}\mathrm{S}/^{34}\mathrm{S}$ exhibit positive radial gradients characteristic of primary-to-secondary isotope ratios (increasing towards the outer disc owing to the lower production of secondary isotopes in less chemically evolved regions), $^{14}\mathrm{N}/^{15}\mathrm{N}$ is predicted to increase towards the Galactic centre (steeply in FR\,CN1 and more mildly in MN\,CN2), reflecting the enhanced production of secondary $^{14}\mathrm{N}$ by metal-rich stars in the inner Galaxy. However, this inner-disc rise in $^{14}\mathrm{N}/^{15}\mathrm{N}$ is at odds with the observational trend reported by \citet{colzi2022}, who found an increase in the ratio with Galactocentric distance up to $\sim 11~\mathrm{kpc}$, followed by a decrease towards the outer Galaxy. This discrepancy may reflect shortcomings in the adopted nucleosynthesis prescriptions, particularly the contribution of novae to $^{15}\mathrm{N}$ production, but could also be partly related to local chemical fractionation effects affecting the molecular isotopologues used to derive the isotopic ratios. More generally, this highlights a fundamental challenge when comparing GCE predictions with molecular cloud observations: molecular isotopologue ratios do not necessarily trace the underlying elemental isotopic composition, as they can be altered by isotope-selective chemical processes. Properly quantifying these fractionation effects is therefore essential to recover the intrinsic isotopic ratios and reliably constrain GCE \citep{colzi2026}. We further compare our models with the combined quantity $(^{14}\mathrm{N}/^{15}\mathrm{N})\times(^{13}\mathrm{C}/^{12}\mathrm{C})$. Since observational estimates of $^{14}\mathrm{N}/^{15}\mathrm{N}$ often employ the double-isotope method---multiplying optically thin column-density ratios such as $\mathrm{H}^{13}\mathrm{CN}/\mathrm{HC}^{15}\mathrm{N}$ or $\mathrm{HN}^{13}\mathrm{C}/\mathrm{H}^{15}\mathrm{NC}$ by an assumed $^{12}\mathrm{C}/^{13}\mathrm{C}$ value \citep[e.g.][]{colzi2018,colzi2022}---this product allows a direct comparison with observed isotopologue ratios without introducing an explicit dependence on the Galactic $^{12}\mathrm{C}/^{13}\mathrm{C}$ gradient. Notably, model MN\,CN2 closely reproduces the observed trend across the full radial range, whereas model FR\,CN1 overpredicts the ratio in the inner disc ($R_{\mathrm{GC}} \lesssim 8~\mathrm{kpc}$). Potential molecule-dependent isotopic fractionation effects, however, should still be kept in mind.

Finally, it should be emphasised that the classical GCE approach used here does not account for dynamical processes such as radial gas inflows and active stellar radial migration. Both mechanisms are known to shape the evolution of Galactic gradients; in particular, radial gas inflows generally steepen gas-phase abundance profiles over time, whereas stellar radial migration tends to flatten stellar gradients with increasing age. A thorough investigation incorporating radial gas flows and active stellar migration will be presented in the forthcoming Paper~II of the SPONGE series (Zou et al., in prep.). The theoretical gradients presented in Fig.~\ref{fig:grad} serve primarily as a benchmark for the nuclear yields adopted in this study.

\section{Discussion and conclusions}
\label{sec:disc}

In this first paper of the SPONGE series, we have investigated the evolution of CNOPS elements and their isotopic ratios in the Milky Way by confronting classical (homogeneous) GCE models with a comprehensive observational dataset compiled from the literature. Assembled to span as wide a range of metallicities and Galactocentric distances as possible, this dataset constrains our theoretical predictions against both the temporal and spatial enrichment histories of these key elements.

Our main conclusions are summarised as follows:
\begin{itemize}
\item Baseline models FR\,CN1 and MN\,CN2, which incorporate nucleosynthesis from single stars and binary systems (SNe~Ia and novae), satisfactorily reproduce the average observed trends across the in-situ halo, thick disc, and thin disc components, alongside the present-day radial abundance gradients of the CNOPS elements. The mild overproduction of $^{12}\mathrm{C}$ predicted by the models, as well as the large star-to-star scatter observed in the metal-poor halo, warrants further investigation.
\item Including nova nucleosynthesis is essential to reproduce the C and N isotopic ratios in the solar neighbourhood, as well as the radial $^{12}\mathrm{C}/^{13}\mathrm{C}$ and $^{14}\mathrm{N}/^{15}\mathrm{N}$ gradients across the disc. While AGB and massive stars account for the bulk of $^{13}\mathrm{C}$ and $^{14}\mathrm{N}$, novae supply the $^{15}\mathrm{N}$ needed to match both the protosolar composition and local molecular cloud measurements without overproducing primary CNO species. The lowest $^{12}\mathrm{C}/^{13}\mathrm{C}$ ratios observed in some metal-poor Galactic stars remain unexplained.
\item Models FR\,CN1 and MN\,CN2 naturally produce negative $[\mathrm{El}/\mathrm{H}]$ radial abundance gradients (for $\mathrm{El} = \mathrm{C}, \mathrm{N}, \mathrm{O}, \mathrm{P}$, and $\mathrm{S}$) and positive radial gradients for $^{12}\mathrm{C}/^{13}\mathrm{C}$, $^{16}\mathrm{O}/^{18}\mathrm{O}$, and $^{32}\mathrm{S}/^{34}\mathrm{S}$ across $R_{\mathrm{GC}} = 4$--$18~\mathrm{kpc}$. These spatial trends mirror the inside-out assembly of the Galactic disc, where higher star formation rates in the inner regions lead to greater cumulative nucleosynthetic processing. The nitrogen isotopic ratio is a notable exception: because $^{14}\mathrm{N}$ behaves predominantly as a secondary isotope at disc metallicities while $^{15}\mathrm{N}$ receives a substantial delayed contribution from novae, $^{14}\mathrm{N}/^{15}\mathrm{N}$ is predicted to increase towards the Galactic centre. A mild increase of the ratio is in agreement with the combined quantity $(^{14}\mathrm{N}/^{15}\mathrm{N})\times(^{13}\mathrm{C}/^{12}\mathrm{C})$ that is directly measured in star-forming regions.
\item Within our current framework, the predicted $^{32}\mathrm{S}/^{34}\mathrm{S}$ ratios along the disc lie systematically below the observed values owing to an overproduction of $^{34}\mathrm{S}$ at disc metallicities in the adopted massive-star yields, the origin of which warrants further investigation by stellar evolution modellers.
\end{itemize}

While the baseline models presented here provide an essential benchmark for validating nuclear yield sets and nova prescriptions, they inherently neglect dynamical transport processes such as radial gas inflows and active stellar radial migration, which are expected to reshape Galactic abundance profiles over time. In Paper~II of the SPONGE series (Zou et al., in prep.), we will self-consistently incorporate these mechanisms to investigate how dynamical evolution shapes the CNOPS chemical landscape across the Milky Way disc over cosmic time.

   \begin{acknowledgements}
    We gratefully acknowledge the financial and logistical assistance provided by the Jules Janssen Conference Centre, supported by the Scientific Council of the Observatoire de Paris. This research was partially funded by the Italian National Institute for Astrophysics (INAF) through the program Finanziamento della Ricerca Fondamentale 2022-23, projects \emph{“An in-depth theoretical study of CNO element evolution in galaxies”}, Fu.~Ob.~1.05.12.06.08 (PI: D.~Romano) and \emph{“Open Clusters and stellar structures in the local Galactic disk”}, Fu.~Ob.~1.05.23.05.19 (PI: A.~Vallenari). Y.~P.~Z. is supported by the China Scholarship Council (CSC) under Grant No.~202408440235. L.~M. gratefully acknowledges support from ANID–FONDECYT Regular Project No. 1251809. L.~C. received the support of a fellowship from the “la Caixa” Foundation (ID 100010434). The fellowship code is LCF/BQ/PR25/12110012. L.~C. acknowledges support from the grant PID2022-136814NB-I00 funded by the Spanish Ministry of Science, Innovation and Universities/State Agency of Research MICIU/AEI/10.13039/501100011033 and by ERDF, UE. E.~S. thanks INAF for the Mini Grant \emph{“LEGARE: Linking the chemical Evolution of Galactic discs AcRoss diversE scales: from the thin disc to the nuclear stellar disc”}, Fu.~Ob.~1.05.24.07.02 (PI: E.~Spitoni). This work has made use of data from the European Space Agency (ESA) mission \textit{Gaia} (\url{https://www.cosmos.esa.int/gaia}), processed by the \textit{Gaia} Data Processing and Analysis Consortium (DPAC, \url{https://www.cosmos.esa.int/web/gaia/dpac/consortium}). Funding for the DPAC has been provided by national institutions, in particular the institutions participating in the \textit{Gaia} Multilateral Agreement. This research has made use of the SIMBAD database, operated at CDS, Strasbourg, France. The authors utilized Gemini to assist with language polishing of the manuscript.
\end{acknowledgements}

\bibliographystyle{aa} 
\bibliography{SPONGE-I}

\begin{appendix}
\appendix

\section{Basic model assumptions and mathematical formulation}
\label{app:basic}

The Galaxy is modelled as a series of independent, concentric cylindrical shells, each with a width of $2\,\mathrm{kpc}$ and a height of $1\,\mathrm{kpc}$. The mass accretion rate per unit area during the two-infall process is expressed as:
\begin{equation}
\dot{\sigma}_{\mathrm{in}}(R, t) = A(R) \, \mathrm{e}^{-t/\tau_1} + \vartheta(t - t_{\mathrm{max}}) \, B(R) \, \mathrm{e}^{-(t-t_{\mathrm{max}})/\tau_2(R)},
\end{equation}
where $\vartheta$ is the Heaviside step function, and the parameters $\tau_1$ and $\tau_2(R)$ denote the e-folding timescales governing gas accretion onto the in situ halo/thick disc and the thin disc, respectively. We fix the first timescale to $\tau_1 = 1\,\mathrm{Gyr}$, whereas the thin disc timescale $\tau_2(R)$ varies with radius to replicate an inside-out growth scenario \citep{larson1976,matteucci1989}, according to the relation \citep{romano2000}:
\begin{equation}
\tau_2(R) = \left( 1.033 \, \frac{R}{\mathrm{kpc}} - 1.267 \right) \mathrm{Gyr}.
\end{equation}
The parameter $t_{\mathrm{max}}$ establishes the time of maximum gas deposition onto the thin disc. Updating earlier formulations with $t_{\mathrm{max}} = 1\,\mathrm{Gyr}$ \citep[e.g.][]{chiappini1997,chiappini2001,romano2010}, we set $t_{\mathrm{max}} = 3.25\,\mathrm{Gyr}$ based on combined elemental abundance and asteroseismic age constraints \citep{spitoni2019,spitoni2024,nissen2020,palla2020,molero2023}. The infalling gas in both episodes is assumed to have a primordial chemical composition set by Big Bang nucleosynthesis \citep{pitrou2021}. The normalization coefficients $A(R)$ and $B(R)$ are analytically determined by fitting the present-day radial profiles of the total surface mass densities in both disc components. Assuming exponential structural configurations, these profiles are defined as:
\begin{equation}
\sigma_{\mathrm{thick}}(R) = \sigma_{0, \mathrm{thick}} \, \mathrm{e}^{-R/2.3},
\end{equation}
with $\sigma_{0, \mathrm{thick}} = 227.0\,M_\odot\,\mathrm{pc}^{-2}$, and
\begin{equation}
\sigma_{\mathrm{thin}}(R) = \sigma_{0, \mathrm{thin}} \, \mathrm{e}^{-R/3.5},
\end{equation}
with $\sigma_{0, \mathrm{thin}} = 393.3\,M_\odot\,\mathrm{pc}^{-2}$. With this choice, we enforce a total baryonic mass surface density of $47\,M_\odot\,\mathrm{pc}^{-2}$ for the solar neighbourhood, in agreement with the dynamical determination by \citet{mckee2015} of $47.1 \pm 3.4\,M_\odot\,\mathrm{pc}^{-2}$. Furthermore, we obtain a relative contribution of the thick disc component to the total density of about 15 per cent, consistent with structural analyses by \citet{blandhawthorn2016} and \citet{mackereth2017}. The predicted present-day local surface densities for visible stars ($34\,M_\odot\,\mathrm{pc}^{-2}$) and stellar remnants ($5\,M_\odot\,\mathrm{pc}^{-2}$) align well with empirical estimates of the local surface densities of main-sequence and giant stars ($27.0 \pm 2.7\,M_\odot\,\mathrm{pc}^{-2}$) and white dwarfs ($4.9 \pm 0.6\,M_\odot\,\mathrm{pc}^{-2}$) compiled by \citet{mckee2015}.

The star formation process is parameterized via a modified Kennicutt-Schmidt law relating the star formation rate (SFR) to the total and gas surface densities:
\begin{equation}
\psi(R, t) = \nu(R) \, \sigma(R, t)^{k-2} \, \sigma_{\mathrm{gas}}(R, t)^k,
\end{equation}
where the power-law exponent is set to $k = 1.5$. The coefficient $\nu(R)$ denotes the star formation efficiency (expressed in $\mathrm{Gyr}^{-1}$), which varies as a function of Galactocentric radius $R$ following the prescriptions detailed in table~3 of \citet{palla2020} during the thin-disc phase. During the halo/thick-disc phase, it is set to $2\,\mathrm{Gyr}^{-1}$ for all radii. For the distribution of stellar masses at birth, we adopt a space- and time-invariant initial mass function (IMF) from \citet{kroupa1993}. This formulation accounts for integrated star formation across the extended local volume \citep[see][]{jerabkova2025}, making it the appropriate choice for GCE models that evaluate average properties smoothed on kiloparsec scales.

The surface density evolution of element $i$ is calculated by solving the integro-differential equation
\begin{equation}
\dot{\sigma}_i(R, t) = -\psi(R, t) \, X_i(R, t) + \dot{\sigma}_{\mathrm{in}}(R, t) \, X_{i, \mathrm{P}} + \dot{\mathcal{R}}_i(R, t),
\label{eq:intdiff}
\end{equation}
where $\sigma_i(R, t) = \sigma_{\mathrm{gas}}(R, t) \, X_i(R, t)$ is the gas surface density in the form of element $i$, $X_i(R, t)$ is the abundance by mass of element $i$ at radius $R$ and time $t$, $X_{i, \mathrm{P}}$ is its primordial mass fraction, and $\dot{\mathcal{R}}_i(R, t)$ represents the stellar return rate of element $i$ per unit area \citep[see e.g.][]{matteucci2021}.

Our models span Galactocentric radii in the range $4$--$18\,\mathrm{kpc}$. The region within $4\,\mathrm{kpc}$ is omitted due to its complex structure \citep[e.g.][]{bovy2019,queiroz2021,eilers2022}. On the other hand, the applicability of our model to the outermost Galactic regions is limited by the abrupt drop in gas volume density and SFR, as well as the break in the stellar density profile beyond $12\,\mathrm{kpc}$ \citep[e.g.][]{fiteni2026}. Finally, radial gas flows and active stellar migration are not implemented in the current models; a thorough discussion of their effects on GCE model predictions is deferred to Paper~II in this series (Zou et al., in prep.).

%
%
%

\section{Kinematic analysis}
\label{app:kin}

To homogeneously analyse the kinematics of our heterogeneous stellar sample, we follow the approach first outlined in \citet[][see also \citealt{bonifacio2024}]{bonifacio2021}. Briefly, we retrieve from the \textit{Gaia} DR3 \citep{gaia2023} database the stars' coordinates, proper motions, parallaxes, radial velocities, their associated uncertainties, and the coefficients of the correlation matrix. Stellar distances are derived by inverting the parallaxes, which are first zero-point corrected following the prescription of \citet{lindegren2021}. For the stars present in the \textit{Gaia} database, kinematic quantities are then calculated by feeding these parameters (coordinates, distances, proper motions, and radial velocities) into the \texttt{galpy} code \citep{bovy2015}. We employ the standard \texttt{MWPotential2014} potential and adopt $8~\mathrm{kpc}$ and $220~\mathrm{km\,s^{-1}}$ as the solar Galactocentric distance and the circular velocity at the solar radius, respectively \citep[see, e.g.,][]{bovy2012}. We further adopt the \citet{schonrich2010} solar peculiar motions.

To estimate the uncertainties on the calculated parameters, we use the \texttt{pyia} code \citep{pricewhelan2018} to extract realisations of the input parameters from a multivariate Gaussian, considering the parameter uncertainties as dispersions and taking into account the covariance matrix. For each star, we repeat the \texttt{galpy} analysis for 1000 realisations of the input parameters and adopt the standard deviations of the resulting distributions as uncertainties on the calculated parameters.

The above analysis is performed only on stars present in the \textit{Gaia} DR3 database that satisfy the following restrictions on specific parameters: \texttt{astrometric\_params\_solved}~$> 3$, \texttt{parallax\_over\_error}~$> 3$, and \texttt{rv\_method\_used}~$\neq 0$. In our case, this corresponds to selecting stars with measured parallaxes and radial velocities whose parallax uncertainty is at least three times smaller than the parallax itself.

    \begin{figure*}
    \centering
    \includegraphics[width=0.49\textwidth]{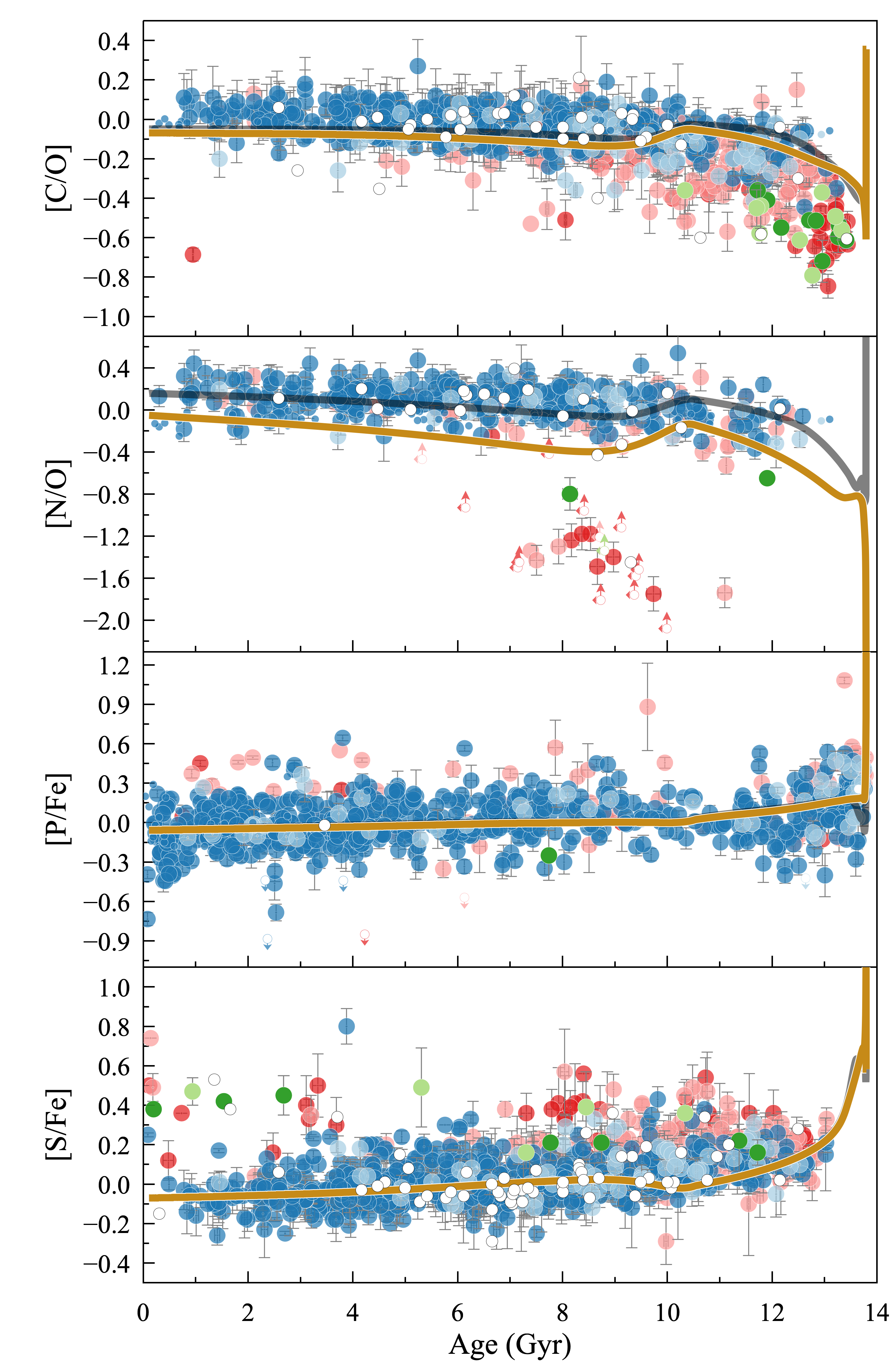}
    \includegraphics[width=0.49\textwidth]{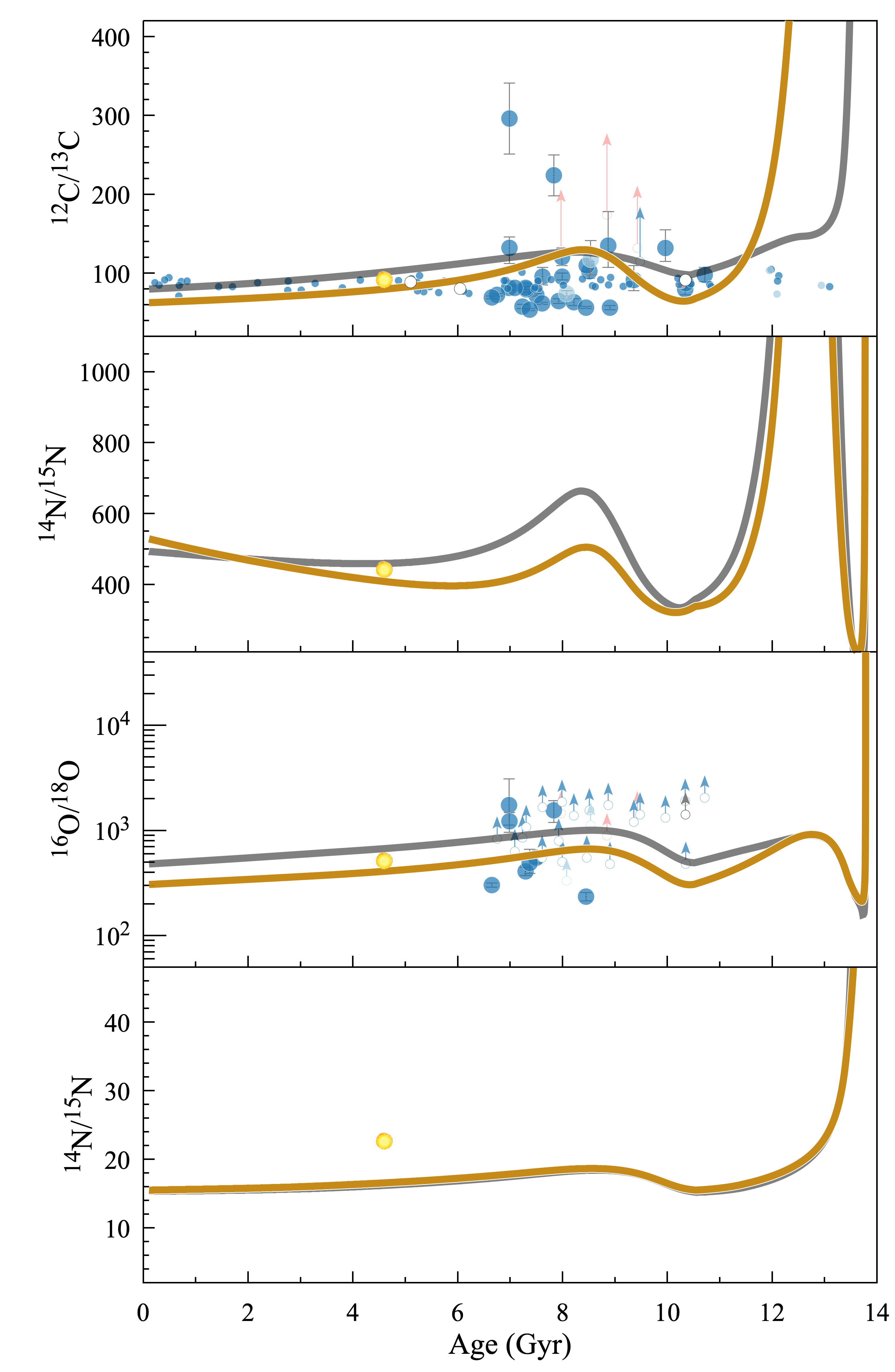}
    \caption{Evolution of $[\mathrm{C}/\mathrm{O}]$, $[\mathrm{N}/\mathrm{O}]$, $[\mathrm{P}/\mathrm{Fe}]$, and $[\mathrm{S}/\mathrm{Fe}]$ \textit{(left panels)}, and $^{12}\mathrm{C}/^{13}\mathrm{C}$, $^{14}\mathrm{N}/^{15}\mathrm{N}$, $^{16}\mathrm{O}/^{18}\mathrm{O}$, and $^{32}\mathrm{S}/^{34}\mathrm{S}$ \textit{(right panels)} as a function of stellar age for stars in the solar neighbourhood (see Sect.~\ref{sec:data_SN}). Observational data are colour-coded according to their kinematic classification (see Appendix~\ref{app:kin}): in-situ halo (red circles), thick disc (pink circles), thin disc (blue circles), transition population (light blue circles), Gaia-Enceladus (dark green circles), and Sequoia (light green circles). Unclassified stars are shown as white circles. Solar twins and open clusters are shown as smaller filled circles. Open circles with arrows indicate upper or lower limits, and yellow circles in the right panels mark the (proto)solar ratios. Solid lines represent the temporal tracks predicted by the baseline GCE models: FR\,CN1 (dark ochre) and MN\,CN2 (grey).}
    \label{fig:age}
    \end{figure*}
    
For a minority of the stars, \textit{Gaia} radial velocities (RVs) were not available, but their RVs were known in SIMBAD, and we adopted these values together with the \textit{Gaia} astrometric parameters. For a few stars for which neither \textit{Gaia} astrometry nor RVs were available, we adopted \textit{Hipparcos} parameters \citep{hipparcos2007} together with RVs from SIMBAD.

Stars are classified as belonging to the halo, the thin or thick disc, or as `transition' objects between the two following the criteria defined in \citet{bensby2014}. We further classify stars as likely Gaia-Enceladus or Sequoia members following the criteria of \citet{feuillet2021}. Notice that this definition of Sequoia includes both a highly retrograde, high-energy component and a mildly retrograde, low-energy component. Sometimes, only the former is referred to as Sequoia, while the latter is associated with the Thamnos accretion event \citep[see, e.g.,][]{koppelman2019}.

\section{Stellar ages}
\label{app:ages}

Traditionally, chemical evolution models have been presented using trends of abundance ratios as a function of the abundance of a reference element. The underlying idea is that the abundance of an element, such as oxygen or iron, is a proxy for time. This choice was a necessary one, since it was difficult, if not impossible, to obtain meaningful ages for field stars. Since reliable parallaxes from the \textit{Gaia} mission \citep{gaia2016,gaia2023} became available, the situation has changed, and from the absolute luminosities and effective temperatures of the stars it is possible to estimate ages by comparison with theoretical evolutionary tracks or isochrones. One aspect that has become apparent is that there is not a single age--metallicity relation (AMR) in the Galaxy, but rather a superposition of several AMRs due to mergers and the distinct evolution of the various Galactic components \citep[see, e.g.,][]{bonifacio2026}.

In this appendix, we explore the comparison of the predictions of our GCE models with observations, using stellar age as the time axis. Since the literature data used to explore the models in the chemical planes, we decided to use evolutionary tracks in the $\log(T_{\mathrm{eff}})$--$\log g$ plane. Some of the papers from which we took abundances used \textit{Gaia} parallaxes to derive $\log g$, though not all of them. Nevertheless, by using these parameters, we derive ages that are coherent with the adopted chemical abundances. Within the framework of the SPONGE project, future papers in this series will present homogeneous abundances and consistent evolutionary ages derived from proprietary observations (either already in hand or secured through approved proposals). In particular, targets have been specifically selected among subgiant stars with parallax-based surface gravities to guarantee robust, high-precision age dating.

To derive the ages of the stars in the combined literature sample (see Sect.~\ref{sec:data_SN}), we used SPInS \citep[Stellar Parameters Inferred Systematically,][]{lebreton2020,reese2020}, a public Python tool based on a Bayesian approach. We adopted the theoretical BaSTI stellar evolutionary tracks, either solar-scaled \citep{hidalgo2018} or $\alpha$-enhanced \citep{pietrinferni2021}, as appropriate, selecting tracks that include overshooting and diffusion. We also adopted a flat prior on age between $0$ and $13.8~\mathrm{Gyr}$ in order to constrain the derived ages to be younger than the age of the Universe \citep{planck2020}.

In Fig.~\ref{fig:age}, the $[\mathrm{C}/\mathrm{O}]$, $[\mathrm{N}/\mathrm{O}]$, $[\mathrm{P}/\mathrm{Fe}]$, and $[\mathrm{S}/\mathrm{Fe}]$ abundance ratios predicted by our baseline models FR\,CN1 and MN\,CN2 as functions of age are compared to observations. Stars in the combined literature sample are colour-coded according to their Galactic component membership (see Appendix~\ref{app:kin} and the caption of Fig.~\ref{fig:novae}). The predicted trends align well with the evolution of the thin disc, while they lie above the data for the in-situ halo, thick disc, and accreted components in the $[\mathrm{C}/\mathrm{O}]$ and $[\mathrm{N}/\mathrm{O}]$ versus age planes, and below the data in the $[\mathrm{S}/\mathrm{Fe}]$ plane. This is suggestive of a parallel evolution, rather than the sequential one postulated here. Only by exploiting stellar age information is it possible to draw such a conclusion, which otherwise remains hidden in the classical $[\mathrm{El}/\mathrm{M}]$ versus $[\mathrm{M}/\mathrm{H}]$ diagnostic diagrams (where $\mathrm{M} = \mathrm{O, Fe}$).

\end{appendix}

\end{document}